\documentclass[10pt,twocolumn,twoside]{IEEEtran}

\usepackage{setspace,multirow,tikz}
\usetikzlibrary{arrows,automata}

\usepackage[T1]{fontenc}
\usepackage[latin9]{inputenc}
\usepackage{amsthm}
\usepackage{amsmath}
\usepackage{graphicx}
\usepackage{amssymb}
\usepackage{url}
\usepackage{verbatim}

\theoremstyle{plain}

\theoremstyle{plain}

\usepackage{cite}\usepackage{amsfonts}\usepackage{times}\usepackage{bm}
\usepackage{amsthm}
\usepackage{subfigure}
\theoremstyle{definition}

\newtheorem*{lemma*}{Lemma}

\theoremstyle{remark}
\newtheorem{remark}{Remark}

\definecolor{orange}{RGB}{34,139,34}

\begin{document}
\title{Queuing models for abstracting interactions in Bacterial communities}
\author{
Nicol\`{o}~Michelusi,~James~Boedicker,~Mohamed~Y.~El-Naggar~and~Urbashi~Mitra
\thanks{N. Michelusi is with the Department of Electrical and Computer Engineering at Purdue University, West Lafayette, USA;
 U. Mitra is with the Ming Hsieh Department of Electrical Engineering, University of Southern California, Los Angeles, USA;
M. Y. El-Naggar and J. Boedicker 
are with the Departments of Physics, Biological Sciences, and Chemistry,
University of Southern California, Los Angeles, USA; emails:
\tt{michelus@purdue.edu,\{boedicke,mnaggar,ubli\}@usc.edu}
}
\thanks{N. Michelusi and U. Mitra acknowledge support from one or all of these grants:  
ONR N00014-09-1-0700, CCF-0917343, CCF-1117896, CNS-1213128, AFOSR FA9550-12-1-0215, and DOT CA-26-7084-00.
J. Boedicker acknowledges support from ONR award N00014-15-1-2573.
M. Y. El-Naggar acknowledges support from NASA Cooperative Agreement NNA13AA92A and PECASE award FA955014-1-0294 from the Air Force Office
of Scientific Research.}
\vspace{-5mm}
}

\maketitle

\begin{abstract}
Microbial communities play a significant role in bioremediation, plant growth, human and animal digestion, global elemental cycles including the carbon-cycle, and water treatment. They are also posed to be the engines of renewable energy via microbial fuel cells which can reverse the process of electrosynthesis. Microbial communication regulates many virulence mechanisms used by bacteria. Thus, it is of fundamental importance to understand interactions in microbial communities and to develop predictive tools that help control them, in order to aid the design of systems  exploiting bacterial capabilities. This position paper explores how abstractions from communications, networking and information theory can play a role in understanding and modeling bacterial interactions. In particular, two forms of interactions in  bacterial systems will be examined: {\em electron transfer} and {\em quorum sensing}. While the diffusion of chemical signals has been heavily studied, electron transfer occurring in living cells and its role in cell-cell interaction is less understood. Recent experimental observations open up new frontiers in the design of microbial systems based on electron transfer, which may coexist with the more well-known interaction strategies based on molecular diffusion. 
 In quorum sensing, the concentration of certain signature chemical compounds emitted by the bacteria is used to estimate the bacterial population size, so as to activate collective behaviors. 
In this position paper, queuing models for electron transfer are summarized and adapted to provide new models for quorum sensing.
These models 
 are stochastic, and thus capture the inherent randomness exhibited by cell colonies in nature.
 It is shown that queuing models allow the characterization of the
state of a single cell as a function of interactions with other cells and
the environment, thus enabling the construction of an information theoretic framework,
while being amenable to complexity reduction using methods based on statistical physics and wireless network design.
\end{abstract}

\vspace{-3mm}
\section{Introduction}
\label{intro}
Bacteria constitute some of the earliest life
forms on Earth, existing anywhere from four to three billion years ago \cite{Kaufman}. 
Bacteria are unicellular organisms that lack 
a nucleus and rarely harbor membrane-bound organelles.
 While bacteria
have certain elements in common (such as being unicellular), their numbers
and variety are vast, with an aggregate biomass larger than that of all animals and plants combined
\cite{hogan2010bacteria}. 
 Bacteria exist all over the planet, in and on living creatures,
underground, and underwater. They can survive in Antarctic lakes and in
hot springs, showing tremendous robustness and tenacity.
 It is theorized that bacteria enabled changes in the Earth's environment
leading to atmospheric oxygen as well as being the precursors to more complex
organisms.
Microbial communities play a significant
role in bioremediation, plant growth, human and animal digestion, global elemental cycles including the carbon-cycle, and water treatment~\cite{KJB}.
  Despite their simplicity
as unicellular organisms, the operations and interactions of bacterial communities are
not fully understood.
Elucidating how bacteria interact with each other and with their environment 
is of fundamental importance in order to fully exploit their potential.
To this end, realistic tools for the modeling and prediction of these systems are needed

In this position paper, we explore how queuing models can play a role in understanding and modeling bacterial
interactions.
Indeed, \emph{molecules}, \emph{electrons} and \emph{cells} are countable units. It is thus natural to represent them as "quanta", and their amount as the state of a "queue" in which these quanta are collected
and from which they are dropped.
Queue evolution then depends on the interaction of these elements with each other and with the surrounding environment. The advantage of a queuing model compared, for instance, to continuous models based on ordinary differential equations, is their amenability to capture randomness and interactions in 
small bacterial systems, such as the quorum sensing system studied in \cite{Boedicker}, where the discrete nature of these interactions is predominant.
 In particular, we shall examine two forms of interactions:

\paragraph{Electron Transfer\cite{Reguera,Pfeffer}}
While chemical and molecular diffusion are widely investigated~\cite{nakano2013molecular,mian2011communication}, electron transfer
 is emerging as an exciting strategy by which bacteria exchange
nutrients and potentially information. Each cell relies on a continuous flow of electrons from an electron donor to an electron acceptor through
the cell's electron transport chain to produce energy in the form of
the molecule adenosine triphosphate (ATP), and to sustain its vital operations
and functions. This strategy, known as \emph{oxidative phosphorylation}, is employed
by all respiratory microorganisms. While the importance of biological electron transport
 is well-known for individual
cells, the past decade has also brought about remarkable discoveries of multicellular
microbial communities that \emph{transfer} electrons between cells and across much
larger length scales than previously thought \cite{Naggar},
spanning from molecular assemblies known as \emph{bacterial nanowires}, to
entire macroscopic architectures, such as biofilms and multicellular bacterial
cables \cite{Reguera,Pfeffer}. 
Electron transfer has been observed
in nature \cite{Pfeffer} and in colonies cultured in the laboratory \cite{Kato}.
  This
multicellular interaction is typically initiated by a lack of either 
electron donor or acceptor, which in turn triggers gene expression.
 The overall goal in these extreme conditions is for the colony to survive
despite deprivation, by relaying electrons from the donor to the acceptor to support the electron transport chain in each cell.
The survival of the whole system relies on this division of labor, with the
intermediate cells operating as {\em relays} of electrons to coordinate this
collective response to the spatial separation of electron donor and acceptor.

These observations open up new frontiers in the modeling and design of microbial systems based on electron transfer, which may coexist with the more well-known interaction strategies based on molecular diffusion.
A logical question to ask is whether classical approaches to cascaded
channels \cite{silverman1955binary,simon} are suitable for the study of bacterial
cables, which are indeed {\em multi-hopped} networks \cite{Pirbadian}.
 A straightforward cut-set bound analysis \cite{coverthomas} shows that the
capacity of such a system is achieved by a decode-and-forward strategy. 
However, in microbial systems, there is a stronger interaction between
the bacteria, due to the coupling of the electron signal with the energetic state of the cells via the electron transport chain.
Thus, alternative strategies 
for analysis and optimization must be undertaken,\footnote{Relay systems employing decode and forward for molecular diffusion based communication systems are considered in \cite{einolghozati2013relaying}.}
which take into account the specificity of the interactions between cells in a bacterial cable.
In turn, it is important to understand how to design capacity-achieving
signaling schemes over bacterial cables, which exploit these specific interactions, as a preliminary step towards optimizing fuel cell design (see Sec.~\ref{applic}).
In Sec.~\ref{BG}, we summarize a stochastic queuing model of electron transfer for a single cell,
which links  electron transfer to the energetic state of the cell (\emph{e.g.}, ATP concentration).
This model abstracts the detailed working of a single cell by using interconnected queues to model signal interactions.
In Sec.~\ref{bactcable}, we show how the proposed single-cell model can be extended to larger communities (\emph{e.g.}, cables, biofilms), by allowing  electron transfer  between neighboring cells.
In Sec.~\ref{itcapacity}, based on such a model, we perform a capacity analysis for a bacterial cable, and discuss the design of signaling schemes.

\paragraph{Quorum Sensing \cite{Bassler,Visick15082005,Nealson70}}
As previously noted, biological systems are known to communicate by diffusing chemical signals
in the surrounding medium \cite{nakano2013molecular, mian2011communication}.
In quorum sensing, the concentration of certain signature
chemical compounds emitted by the bacteria is used to
estimate the bacterial population size, so as to regulate collective gene expression.
  However, this simplistic explanation does
not fully capture the complexity and variety of quorum sensing related processes.
In this paper, we provide a model for quorum sensing, informed by our work on electron transfer, that will enable
more sophisticated study and analyses.

While quorum sensing is based on the emission, diffusion and detection of these chemical signatures across the environment in which the cell colony lives,
it differs from recent work endeavoring to model and design transceiver algorithms based on molecular diffusion channels
(see, {\em e.g.}, \cite{kuran2012interference, arjmandi2013diffusion, mosayebi2014receivers, noel2014optimal})
and to analyze their capacity (see, {\em e.g.}, \cite{srinivas2012molecular,li2014capacity, einolghozati2013design,pierobon2013capacity}).
The focus of these works is the design of models and algorithms for engineered systems for future nanomachines to act as transmitters and receivers.  Herein, instead, we
attempt to understand natural systems and their optimization via the control of environmental conditions. 
Rather than looking at the characteristics of the diffusion channel, we consider the system (the cell colony and the surrounding environment) as a whole,
and attempt to model the dynamic interactions between cells and of the cells with the environment.
We abstract the diffusion channel by considering a queueing model to represent it in terms of concentration of signals.
As will be seen in our model depicted in Figs.~\ref{figQSmodel_a} and~\ref{figQSmodel_b},  quorum sensing does not fit well into traditional multi-terminal frameworks such as the multiple access channel   \cite{ shannon1961two, van1971discrete, liao1972multiple, ahlswede1974capacity}, the broadcast channel \cite{ cover1972broadcast,cover1998comments} or two-way communication \cite{shannon1961two}.
In fact, recent work examining capacity questions relating to quorum sensing have considered the binding of molecules and multicellular processes with molecular diffusion \cite{einolghozati2011consensus,einolghozati2013design}.  The models adopted therein do not take the dynamics of the binding processes into consideration as we do here.
\vspace{-3mm}
\subsection{Prior Art on Mathematical Modeling of Bacterial Populations}
\label{sec-system-model}
The modeling of natural populations such as swarms (motile bacteria, swimming
fish, flying birds, or migratory herds of animals) has been consistently
studied over the years.  \emph{Microscopic models} 
model individuals as 
point particles via dynamical equations \cite{Vicsek1995,Vicsek2012,Song2014}. These models have
precision, but become much more complex as the interactions are finely modeled
and do not scale well as the population size increases.  To combat such issues,
the computational approach of simulating a large number of agents with prescribed
interaction rules has also been considered \cite{Vicsek1995,Mina2012}.
These studies duplicate experimental results well, but do not lead to design
methodologies or optimization strategies as the underlying operations are
not considered. In contrast, \emph{macroscopic models} model biological swarm dynamics
via advection-diffusion-reaction  partial differential equations
 \cite{baker2010microscopic,Klapper2010,Song2014,Hammond2013,Friedman2014}.
The partial differential equations are used to describe the evolution of the probability distribution
of group sizes or fraction of individuals with specific characteristics (\emph{e.g.}, motion
orientation \cite{yates2011rsp}). These approaches also suffer from computational
complexity, often relying on numerical solvers, and do not have an eye towards
optimization and control.

From an information theoretic perspective, there have been ongoing efforts to examine how entropy and mutual information\footnote{See \cite{waltermann2011information} for a review of the application of information theory to biology.} can provide insights into 
experimental data \cite{mehta2009information,cheong2011information,perez2011noise}; however these works do not endeavor to model the underlying processes, but rather compute empirical information theoretic measures based on experimental data.  Another key distinction is a lack of an explicit modeling of the dynamics of the constituent systems.

Our modeling approach differs from these mathematical models on the following
aspects: 1) We propose an accurate queueing approach to characterize the
state of a single cell as a function of interactions with other cells and
the environment. Thus, not only this approach can be used to describe the metabolic state
of a cell, but it also enables the construction of an information
theoretic framework;
2) our proposed models are inherently stochastic (rather than deterministic), and thus capture the inherent randomness exhibited by cell colonies in nature;
3) finally, they are amenable to complexity reduction, \emph{e.g}, using methods based on statistical physics~\cite{Mitra}
or wireless network design~\cite{Levorato}. We show an example of application of these principles to the capacity analysis of bacterial cables in Sec.~\ref{itcapacity}.
\vspace{-3mm}
\subsection{Applications}
\label{applic}
The proposed queueing models serve as powerful predictive tools, which will aid the design of systems exploiting bacterial capabilities.
We highlight two relevant applications.
\subsubsection{Microbial Fuel Cells}
Renewable energy technologies
based on bioelectrochemical systems are
now attracting tens of millions of dollars in government and industry funding~\cite{Naggar}. Microbial fuel cells and the essentially reverse
process of microbial electrosynthesis are well known bioelectrochemical technologies;
both use microorganisms to either generate electricity or biofuels in a sustainable
way \cite{rabaey2010microbial}.
 Microbial fuel cells are constructed with microbes
that oxidize diverse organic fuels, including waste products and raw sewage,
while routing the resulting electrons to large-area electrodes where they
are harvested as electricity\cite{logan2009exoelectrogenic}.
 Certain bacterial
strains (\emph{e.g.}, {\em Shewanella oneidensis} \cite{Kim2,logan2009exoelectrogenic}) 
are of particular interest for microbial fuel cells as they can attach to electrodes
and transfer electrons without {\em mediators} which are often toxic, especially
in the high concentrations needed to overcome their diffusion.

\begin{figure}[t]
\centering
\includegraphics[width =.8\linewidth,trim=20 10 15 5,clip=false]{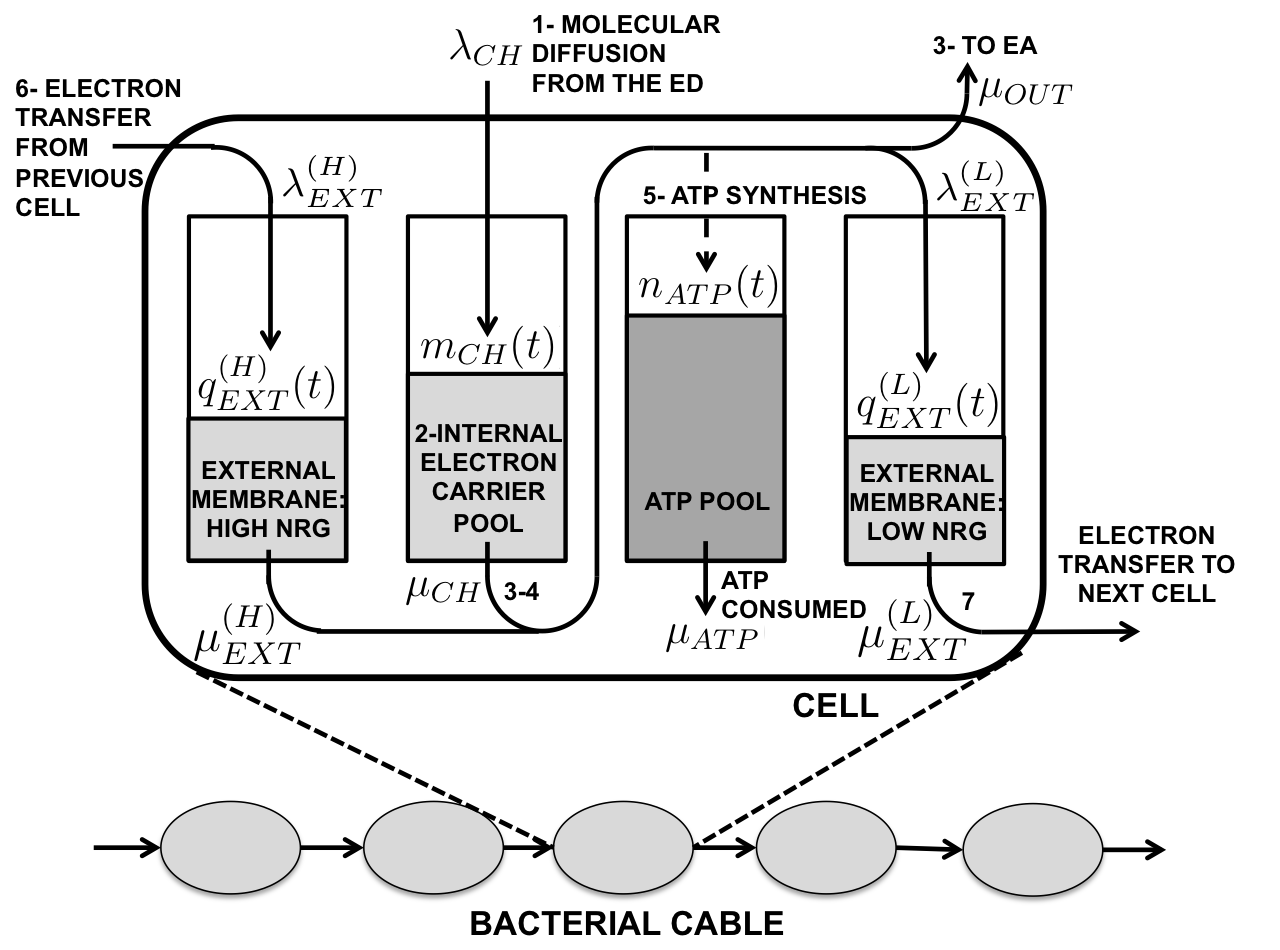}
\caption{Stochastic model of electron transfer within a bacterial cable.}
\label{fig1}
\vspace{-7mm}
\end{figure}

The challenge to realizing microbial fuel cells is understanding the cooperative and anti-cooperative
behaviors of collections of bacteria.  Ideally, as the number of bacteria
increases, the amount of current generated should increase as well.  However,
this is not always observed in many experimental scenarios \cite{mclean2010quantification}.
We posit that the cell-cell interactions necessitated by a multicellular
(\emph{i.e.} biofilm) lifestyle, including electron transfer  across and
through
the biofilms themselves, can limit device performance. In other
words, the biofilm's energy output cannot be casually gleaned from what we
know about the mechanisms relevant to individual or ``disconnected'' microbes.
Optimization of biofilms, and multicellular performance in general, requires
sophisticated models and predictive tools that account for the electronic and nutrient fluxes
within whole communities. In this paper, we propose queuing models to serve this purpose.

\subsubsection{Suppression of Infections}
Quorum sensing processes have been of significant interest since first observed in \emph{Vibrio harveyi} \cite{Nealson70}.  The discovery of quorum sensing in \emph{Pseudomonas aeruginosa}, a human pathogen associated with cystic fibrosis and burn wound infections, demonstrates the medical implications of quorum sensing regulation \cite{Gambello,Singh}.  In \emph{Pseudomonas aeruginosa} 
infections, quorum sensing plays an important role in activating many critical virulence pathways \cite{Erickson,Rumbaugh}.  The genes regulated by quorum sensing enable the bacteria to mount a successful attack or evade the body's defenses.  In these cases, blocking quorum sensing activation may help to prevent or treat microbial-associated diseases \cite{Rasmussen}.  As bacteria build resistance to antibiotics, reducing virulence through inhibition of quorum sensing may be more effective in mitigating disease versus eradicating the bacteria \cite{Allen}. Naturally occurring compounds and enzymes exist which can inhibit quorum sensing \cite{Rasmussen,Kalia}.  However, the complexity of signaling interactions within complex communities of microbes
points to the need for a more
 quantitative understanding of quorum sensing regulation.  Since many different types of bacteria 
use a similar set of signals, and many signals work in conjunction with other autoinducers within the same cell, the potential outcomes of quorum sensing manipulation are not obvious. The combination of predictive model development with quantitative experiments to parametrize and test models are essential to develop successful strategies for virulence reduction. 

This paper is organized as follows. In Secs.~\ref{BG} and \ref{bactcable}, we propose a model of electron transfer for a single cell and a bacterial cable, respectively, 
and present some capacity results.
In Sec.~\ref{QSmodel}, we present a model of quorum sensing in a homogeneous bacterial community.
In Sec.~\ref{simres}, we present some simulation results and experimental data
 and, in Sec.~\ref{extensions}, we discuss extensions of the proposed quorum sensing model. Finally, in Sec.~\ref{conclusions}, we conclude the paper.
\vspace{-3mm}
\section{Single Cell Model of Electron Transfer}
\label{BG}
In this section, we propose a queuing model of  electron transfer in a single cell. We will extend this model
to interconnections of cells in a bacterial cable in Sec.~\ref{bactcable}.
The cell is modeled as a system with an input electron flow coming either
from the electron donor via molecular diffusion, or from a neighboring cell via electron transfer, and an output  flow of electrons leaving either toward the electron acceptor via
molecular diffusion, or toward the next cell in the cable via electron transfer (Fig.~\ref{fig1}).
   Inside the cell, the conventional pathway of electron flow, enabled by
the presence of the electron donor and the acceptor, is as follows (see the numbers in Fig.~\ref{fig1}): 
  electron donors permeate inside the cell via molecular diffusion (1), resulting
in reactions that produce electron-containing carriers (\emph{e.g.}, NADH)
(2),
which are collected in the \emph{internal electron carrier pool}.
 The electron carriers diffusively transfer electrons to the electron transport chain, which are
then discarded by either a soluble and internalized electron acceptor (\emph{e.g.}, molecular
Oxygen) or are transferred through the periplasm to the outer membrane and
deposited on an extracellular electron acceptor (3).
The electron flow through the electron transport chain results in the production of a proton concentration
gradient (proton motive force, \cite{Lane}) across the inner membrane of
the cell (4),
which is utilized by the inner membrane protein \emph{ATP synthase} to produce
ATP, collected in the \emph{ATP pool} and later used by the cell as an energy
source (5).
\vspace{-3mm}
\subsection{Stochastic cell model}
\label{stochmodel}
In this section, we present our proposed cell model.
Each cell incorporates four pools (see  Fig.~\ref{fig1}), each with an associated
state, as a function of time $t$:

\noindent 1) The \emph{internal electron carrier pool}, which contains the electron carrier molecules
(\emph{e.g.}, NADH) produced as a result of electron donor
diffusion throughout the cell membrane and chemical processes occurring inside
the cell,
with state $m_{CH}(t)\in\mathcal M_{CH}\equiv\{0,1,\dots,M_{CH}\}$
representing the number of electrons\footnote{While in the following analysis we assume that one queuing "unit" corresponds to one electron, this can be generalized to the case
where one "unit" corresponds to $N_E$ electrons.}
 bonded to electron carriers in the {internal electron carrier pool},
where $M_{CH}$ is the \emph{electro-chemical storage capacity};
 
\noindent 2) The \emph{ATP pool}, containing the ATP molecules produced via
electron transport chain,
with state $n_{ATP}(t)\in\mathcal N_{AXP}\equiv\{0,1,\dots,N_{AXP}\}$,
representing the number of ATP molecules within the cell,
where $N_{AXP}$ is the ATP capacity of the cell;

\noindent 3) The \emph{external membrane pool}, which involves the extracellular
respiratory pathway of the cell in the outer membrane.
This is further divided into two parts:
a) a \emph{high energy\footnote{Note that the terms \emph{high} and \emph{low} referred to the energy of electrons
 are used here only
in relative terms, \emph{i.e.}, relative to the redox potential at the  cell surface.
In bacterial cables, the redox potential slowly decreases along the cable, thus inducing a net electron flow.}
 external membrane}, which contains high energy
electrons coming from previous cells in the cable;
and
b) a \emph{low energy external membrane}, which collects low energy
electrons from the electron transport chain, before they are transferred
to a neighboring cell. We denote the number of electrons in the high energy and
low energy external membranes as 
$q_{EXT}^{(H)}(t)\in\mathcal Q_{EXT}^{(H)}\equiv \{0,1,\dots,Q_{EXT}^{(H)}\}$,
and $q_{EXT}^{(L)}(t)\in\mathcal Q_{EXT}^{(L)}\equiv \{0,1,\dots,Q_{EXT}^{(L)}\}$, respectively,
where $Q_{EXT}^{(H)}$ and $Q_{EXT}^{(L)}$ are the respective electron ``storage capacities''.
 The \emph{internal cell state} at time $t$ is thus
\begin{align}
\mathbf s_I(t)=\left(m_{CH}(t),n_{ATP}(t),q_{EXT}^{(L)}(t),q_{EXT}^{(H)}(t)\right).
\end{align}  
Cell behavior is also influenced by the concentrations $\sigma_D(t)$ and $\sigma_E(t)$ of the electron donor
and acceptor in the surrounding medium, respectively.
Therefore, we define the \emph{external state} as
$\mathbf s_{E}(t)=(\sigma_{D}(t),\sigma_{A}(t))$.

 \begin{table}[t]
\caption{Description of the queuing abstraction for the electron transfer model.}
\label{tableET}
\begin{center}
\footnotesize
\scalebox{0.88}{
\begin{tabular}{| c | c | c | l |}
 \hline	
\bf Queue & \bf Arrival rate & \bf Service rate & \bf Explanation\\ \hline
$q_{EXT}^{(H)}(t)$ & $\lambda_{EXT}^{(H)}$ & $\mu_{EXT}^{(H)}$ & High energy external membrane  \\ \hline
$q_{EXT}^{(L)}(t)$ & $\lambda_{EXT}^{(L)}$ & $\mu_{EXT}^{(L)}$ & Low energy external membrane  \\ \hline
$m_{CH}(t)$ & $\lambda_{CH}$ & $\mu_{CH}$ & Internal electron carrier pool \\ \hline
$n_{ATP}(t)$ & $\mu_{CH}+\mu_{EXT}^{(H)}$ & $\mu_{ATP}$ & ATP pool \\ \hline
\end{tabular}
}
\end{center}
\vspace{-7mm}
\end{table}

Each pool in this model has a corresponding inflow and outflow of electrons,
each modeled as a Poisson process with rate function of the internal and external state
$(\mathbf s_I(t),\mathbf s_E(t))$. The intensities of these Poisson processes are labeled as in Fig.~\ref{fig1}. For notational simplicity,
the \emph{input} and \emph{output} flows are denoted as $\lambda$ and $\mu$, respectively,
which commonly correspond to  \emph{arrival} and \emph{service} rates in the queueing literature.
A description of the queuing model abstraction is given in Table~\ref{tableET}.
\vspace{-4mm}
\subsection{Model Validation for an Isolated Cell}
\label{isolatedcell}
The experimental investigation of a multi-cellular network of bacteria is very challenging,
 due to the technical difficulties of placing multiple cells next to each other in a controlled way,
 and of performing measurements of electron transfer, ATP and NADH concentrations along the cable \cite{JSACmiche}.
 Therefore, we start by investigating the properties of single, isolated cells, which constitute the building blocks of more complex multicellular systems.

\begin{figure}[t]
\begin{center}
\setlength{\tabcolsep}{1mm}
\begin{tabular}{cc}
\subfigure{\includegraphics*[width=0.5\linewidth,trim=0 5 0 5,clip=true]{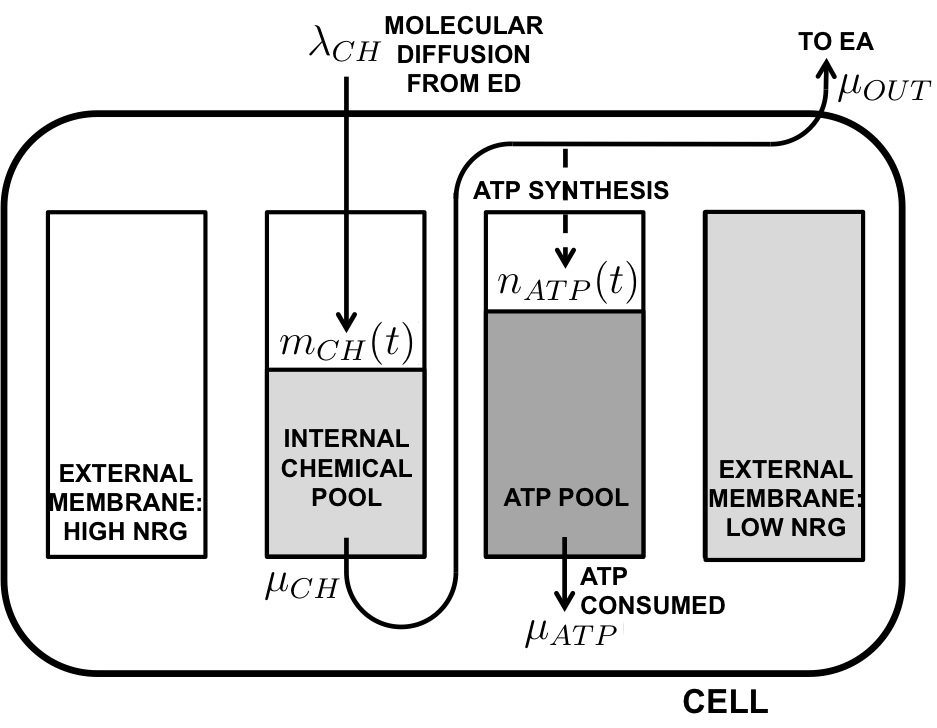}}&
\subfigure{\includegraphics*[width=0.4\linewidth]{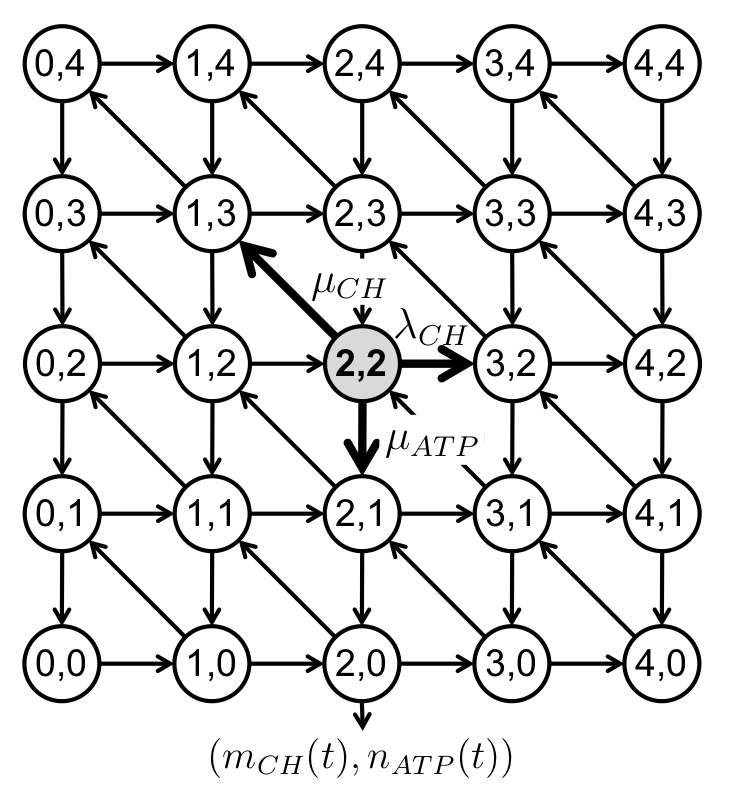}}
\end{tabular}
\caption{
Stochastic model for an isolated cell after the transient phase,
and
Markov chain with the corresponding transitions, for the case where $M_{CH}=N_{AXP}=4$.
The transition rates from state $(2,2)$ are also depicted.}
\label{fig5}
\end{center}
\vspace{-7mm}
\end{figure}

In the case of an isolated cell, multicellular electron transfer does not occur,
hence $\mu_{EXT}^{(L)}=\lambda_{EXT}^{(H)}=0$ (see Fig.~\ref{fig1} for an explanation of the intensities $\mu_{EXT}^{(L)}$ and $\lambda_{EXT}^{(H)}$).
Therefore, after a transient phase during which the 
low and high energy external membranes get
emptied and filled, respectively, the cell's state and the corresponding
Markov chain and state transitions are as depicted in Fig.~\ref{fig5}.
In \cite{JSACmiche},
we have validated this model based on experimental data available in \cite{Ozalp},
using the following parametric model for the rates, inspired by biological
constraints (\emph{e.g.}, Fick's law of diffusion \cite{smith2003foundations}):
\begin{align}\label{flows}
\!\!\!\!\!\begin{array}{l}
\lambda_{CH}(\mathbf s_I(t);\mathbf s_E(t))=\rho\left(1-\frac{m_{CH}(t)}{M_{CH}}\right)\sigma_{D}(t),\\
\mu_{OUT}(\mathbf s_I(t);\mathbf s_E(t))=\mu_{CH}(\mathbf s_I(t);\mathbf s_E(t))\\\qquad\qquad=\zeta\left(1-\frac{n_{ATP}(t)}{N_{AXP}}\right)\sigma_{A}(t),\\
\mu_{ATP}(\mathbf s_I(t);\mathbf s_E(t))=\beta\sigma_{D}(t),
\end{array}\!\!\!\!\!\!\!
\end{align}
 where  $\rho,\zeta,\beta\in\mathbb R_+$ are parameters, estimated
via curve fitting, and $M_{CH}=N_{AXP}=20$.

Initially, cells are starved and the electron donor concentration is $0$. At time $t=80\mathrm{s}$, glucose is added to the cell culture.
Afterwards, the electron donor concentration profile decreases in a staircase fashion from $\sigma_{D}(t)=30\mathrm{[mM]}$
glucose at time $t=80\mathrm{s}$, to
$\sigma_{D}(t)=0\mathrm{[mM]}$ at time $t=1300\mathrm{s}$, and $\sigma_{D}(t)=0\mathrm{[mM]}$
for $t<80\mathrm{s}$,
whereas the electron acceptor concentration (molecular Oxygen) is constant throughout the
experiment, and sufficient to sustain reduction ($\sigma_{A}(t)=~1,\ \forall
t$)~\cite{Ozalp}.
We have designed a parameter estimation algorithm based on maximum likelihood principles, outlined in \cite{JSACmiche}, to fit
the parameter vector $\mathbf x=[\rho,\zeta,\beta]$.

\begin{figure}
\begin{center}
\includegraphics*[width=0.9\linewidth,trim = 0mm 0mm 0mm 0mm,clip=true]{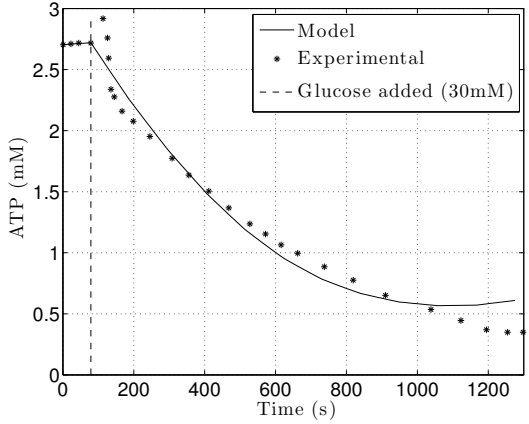}
\caption{Prediction of expected ATP level over time (in $\mathrm{mM}$) and
experimental time-series \cite{JSACmiche}.}
\label{figATP}
\end{center}
\vspace{-7mm}
\end{figure}

Fig.~\ref{figATP} plots the ATP time-series, related to the cell culture, and the expected predicted values
based on our proposed stochastic model. We observe a good fit of the prediction
curves to the experimental ones.
In accordance with \cite{Ozalp}, 
the addition of the electron donor to the suspension of starved cells triggers an increase in ATP and NADH production as well as ATP consumption, as observed experimentally and predicted by our model.
Further discussion can be found in \cite{JSACmiche}.

\vspace{-3mm}
\section{Towards a Model of Bacterial Cables}
\label{bactcable}

In multicellular structures, such as bacterial cables, 
 an additional pathway of electron flow may co-exist,
 termed \emph{intercellular electron transfer}, which involves 
a transfer of electrons
between neighboring cells, as opposed to molecules (electron donor and acceptor) diffusing
through the cell membrane. In this case, one or both of the electron donor and the acceptor
are replaced by neighboring cells in a network of inter-connected cells.
This cooperative strategy creates a multicellular electron transport chain that utilizes intercellular electron transfer
to distribute electrons throughout an entire bacterial network. Electrons
originate from the electron donor localized on one end of the network and terminate to
the electron acceptor on the other end. The collective electron transport through this network
provides energy for all cells involved to sustain their vital operations.

Since typical values for transfer rates between electron carriers (\emph{e.g.}
outer-membrane cytochromes) on the cell exterior are relatively high \cite{Pirbadian},
we assume that $\mu_{EXT}^{(L)}=\lambda_{EXT}^{(H)}=\infty$ when two cells
are connected, so that any electron collected in the low energy external membrane is instantaneously
transferred to the  
high energy external membrane of the neighboring cell in the cable. Therefore, the low energy and high energy external membrane pools
of two neighboring cells can be joined
into a common pool for intercellular electron transfer.
 The \emph{internal state} of cell $i$ at time $t$ is thus
\begin{align}
\mathbf s_I^{(i)}(t)=\left(m_{CH}^{(i)}(t),n_{ATP}^{(i)}(t),q_{EXT}^{(L)(i)}(t),q_{EXT}^{(H)(i)}(t)\right),
\end{align}
and its \emph{external state} is
$\mathbf s_{E}^{(i)}(t){=}(\sigma_{D}^{(i)}(t),\sigma_{A}^{(i)}(t))$.
  The state of the cable of length $N$ is denoted as  $\mathbf s(t){=}(\mathbf
s^{(1)}(t),\dots,\mathbf s^{(N)}(t))$, where $\mathbf
s^{(i)}(t){=}(\mathbf s_I^{(i)}(t),\mathbf s_E^{(i)}(t))$.

The internal state $\mathbf s_I^{(i)}(t)$ is time-varying and stochastic,
and evolves as a consequence of electro/chemical reactions occurring within
the cell, chemical diffusion through the cell membrane, and intercellular electron transfer from the neighboring
cell $i-1$ to cell $i$, and then to the neighboring cell $i+1$.
 The evolution of $\mathbf s_I^{(i)}(t)$ is also influenced by the external
state $\mathbf s_E^{(i)}(t)$ experienced by the cell.
 Let $\mathbf s_I^{(i)}(t)=\left(m_{CH},n_{ATP},q_{EXT}^{(L)},q_{EXT}^{(H)}\right)$
be the state of the $i$th cell at time $t$, and $t^+$ be the time instant
 immediately following $t$.
  We define the following processes which affect the evolution of $\mathbf
s_I^{(i)}(t)$ (see Fig.~\ref{fig1}):
\begin{list}{\labelitemi}{\leftmargin=0.5em}
\item \emph{Electron donor diffusion} through the membrane, joining the internal electron carrier pool with
rate
$\lambda_{CH}(\mathbf s_I^{(i)}(t);\mathbf s_{E}^{(i)}(t))$.
The state becomes
$\mathbf s_I^{(i)}(t^+)=\left(m_{CH}+1,n_{ATP},q_{EXT}^{(L)},q_{EXT}^{(H)}\right)$;
\item \emph{Intercellular electron transfer} from the neighboring cell $i-1$, joining the high energy external membrane
with rate $\lambda_{EXT}^{(L)}(\mathbf s_I^{(i-1)}(t);\mathbf s_E^{(i-1)}(t))$
(in fact, owing to the high transfer rate approximation,
the electron joining the low energy external membrane of cell $i-1$ is immediately transferred to
the high energy external membrane of cell $i$), resulting in
 $\mathbf s_I^{(i)}(t^+)=\left(m_{CH},n_{ATP},q_{EXT}^{(L)},q_{EXT}^{(H)}+1\right)$;
\item \emph{Conventional ATP synthesis}: this process involves the transfer
of one electron from the internal electron carrier pool
to the internal membrane  with rate $\mu_{CH}(\mathbf s_I^{(i)}(t),\mathbf
s_E^{(i)}(t))$, resulting in the synthesis of one unit of ATP.
The electron then leaves the internal membrane
and either follows the \emph{aerobic pathway} (\emph{i.e.}, it is captured
by an electron acceptor), with rate
$\mu_{OUT}(\mathbf s_I^{(i)}(t);\mathbf s_E^{(i)}(t))$, resulting in $ \mathbf
s_I^{(i)}(t^+)= (m_{CH}-1,n_{ATP}+1,q_{EXT}^{(L)},q_{EXT}^{(H)})$,
 or the \emph{anaerobic} one (\emph{i.e.}, it is collected in the low energy external membrane, also
the high energy external membrane of cell $i+1$), with rate $\lambda_{EXT}(\mathbf s_I^{(i)}(t);\mathbf
s_E^{(i)}(t))$,
 so that $\mathbf s_I^{(i)}(t^+)=(m_{CH}-1,n_{ATP}+1,q_{EXT}^{(L)}+1,q_{EXT}^{(H)})$;
  \item \emph{Unconventional ATP synthesis}: this process involves the
transfer of one electron from the 
  high energy external membrane to the internal membrane to synthesize one unit of ATP, with rate
$\mu_{EXT}^{(H)}(\mathbf s_I^{(i)}(t),\mathbf s_E^{(i)}(t))$.
  Afterwards, the electron follows
  either the aerobic pathway with rate $\mu_{OUT}(\mathbf s_I^{(i)}(t);\mathbf
s_E^{(i)}(t))$,
resulting in  $\mathbf s_I^{(i)}(t^+)=(m_{CH},n_{ATP}+1,q_{EXT}^{(L)},q_{EXT}^{(H)}-1)$;
or the anaerobic one with rate $\lambda_{EXT}^{(L)}(\mathbf s_I^{(i)}(t);\mathbf
s_E^{(i)}(t))$,
so that $\mathbf s_I^{(i)}(t^+)=(m_{CH},n_{ATP}+1,q_{EXT}^{(L)}+1,q_{EXT}^{(H)}-1)$;
\item \emph{ATP consumption}, with rate $\mu_{ATP}(\mathbf s_I^{(i)}(t);\mathbf
s_E^{(i)}(t))$,
resulting in $\mathbf s_I^{(i)}(t^+)=(m_{CH},n_{ATP}-1,q_{EXT}^{(L)},q_{EXT}^{(H)})$.
\end{list}
\vspace{-3mm}
\subsection{Information capacity of bacterial cables}
\label{itcapacity}
 In \cite{MicheIT},
 we have studied the information capacity of bacterial cables.
By treating the bacterial cable as a communication medium,
 its capacity represents the maximum amount of information that can be transferred through the cable,
 and thus sets the limit on the rate at which cells can communicate via electron transfer.
 
  Importantly, as shown in \cite{MicheIT} and in the following analysis, communication over a bacterial cable entails two conflicting factors: 
 1) achieving high instantaneous information rate; 2) inducing the bacterial cable to operate in
 information efficient states.
 In fact, the input electron signal affects the energetic state of the cells along the cable via the electron transport chain,
 and thus affects their ability to relay electrons and to transfer information.

Indeed, our analysis reveals that the size of the ATP queues along the cable is extremely important to the health of the cable
and to its capacity to transfer information (or electrons). Thus, with a \emph{greedy} approach (denoted as 
\emph{myopic signaling} and analyzed in Sec. \ref{myopicsignaling}), which ignores these biological constraints,
the overall achievable rate is measurably reduced compared to the optimal capacity achieving scheme, analyzed in Sec. \ref{capacityanalysis}.
Thus, the importance and impact of including the unique constraints and dynamics of bacterial cables is underscored.
 
   \begin{figure}[t]
\centering
\includegraphics[width = .85\linewidth,trim = 10mm 4mm 10mm 9mm,clip=false]{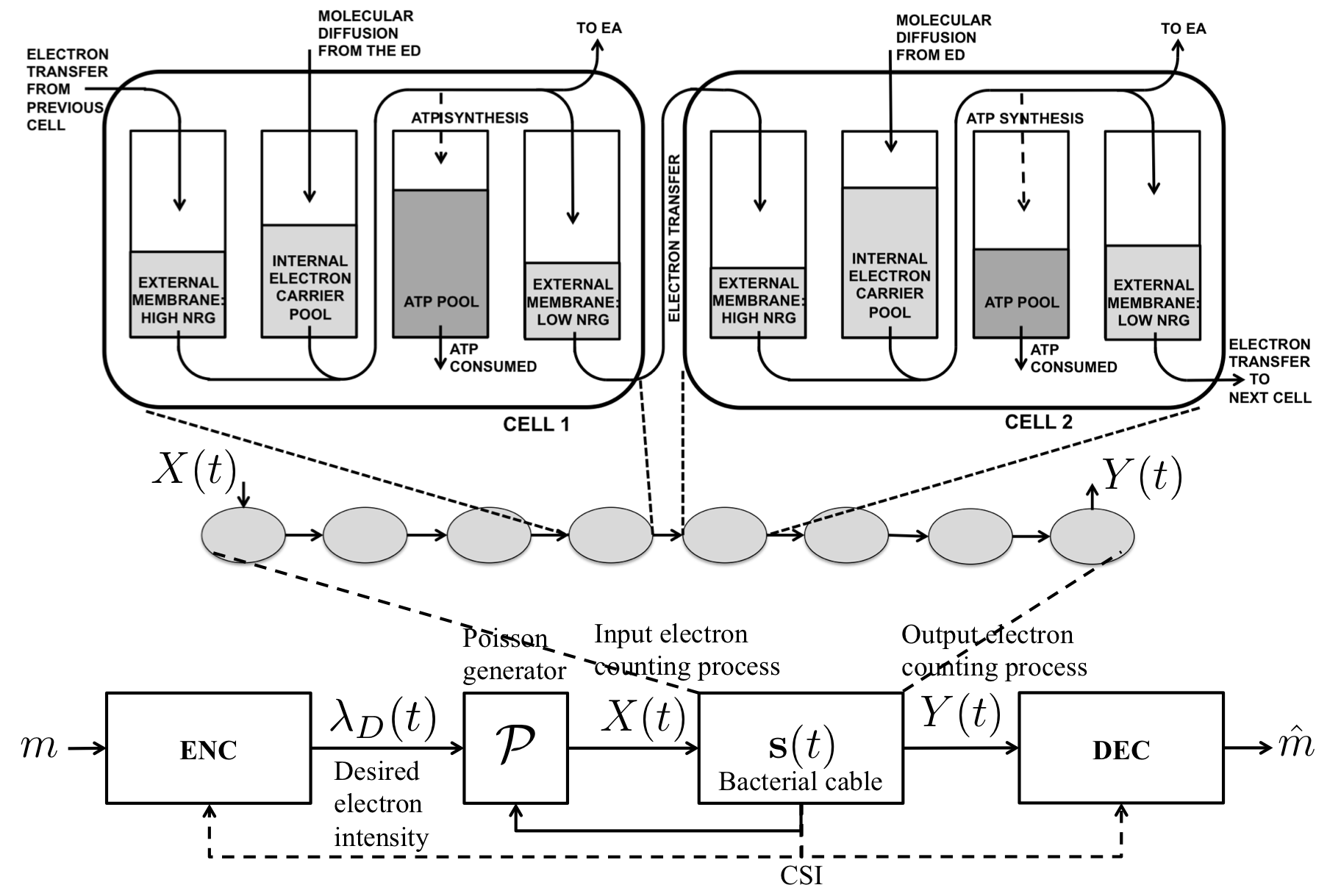}
\caption{Communication system over a bacterial cable. "ENC"=encoder, "DEC"=decoder \cite{MicheIT}.}
\label{baccable}
\vspace{-7mm}
\end{figure}

  \subsubsection{Queuing model}
 The model \cite{JSACmiche} presumes that the channel state is given by the interconnection of the
state of each cell in the cable, leading to high dimensionality.
 As shown in Fig.~\ref{baccable}, the bacterial cable contains $N$ cells.
Letting $\mathcal S_{cell}$
be the state space of the internal state of each cell, then the overall state space  of the cable is
$\mathcal S\equiv\mathcal S_{cell}^N$, which grows exponentially with the bacterial cable length $N$.
Indeed, one of the advantages of our proposed queuing model is its amenability to complexity reduction.
Herein, we propose an abstraction of \cite{JSACmiche} by treating
 the bacterial cable as a black box, which captures only the \emph{global} effects on the 
electron transfer efficiency of the cable, resulting from  the \emph{local} interactions and cells' dynamics presumed by~\cite{JSACmiche}. 
Specifically,
we let  $E(t)$ be the number of electrons carried in the cable, \emph{i.e.}, the sum of the number of electrons carried
in the external membrane of each cell, which participate in the electron transport chain to produce
ATP for the cell.\footnote{Our model in \cite{MicheIT} studies a more general model with leakage and interference.}
The state space for this approximate model is $\mathcal E\equiv\{0,1,\dots, E_{\max}\}$,
where $E_{\max}$ is the electron carrying capacity of the bacterial cable. 
Letting $ E_{\max}^{(cell)}$ be the electron carrying capacity of a single cell,
we have that  $E_{\max}=N\cdot E_{\max}^{(cell)}$, which scales linearly with the cable length, rather than exponentially.

The rationale behind this approximate model is as follows:
when $E(t)$ is large, the large number of electrons in the cable can sustain 
a large ATP production rate, so that the ATP pools of the cells are full and the cable is clogged.
When this happens, the electron relaying capability of the bacterial cable is reduced, 
so that only a portion of the input electrons can be relayed.
On the other hand, when $E(t)$ is small, a weak electron flow occurs along the cable, so that
the ATP pools are almost empty, the cells are energy-deprived, and 
the cable can thus sustain a large input electron flow to recharge the ATP pools.
 We thus define the \emph{clogging function} $\alpha(E(t))\in[0,1]$,
 which represents the fraction of electrons which can be relayed at the input of the cable.
 This model is in contrast to classical relay channels over cascades, where the performance is dictated by the worst link rather than
 by a clogging function \cite{coverthomas}. 

The communication system includes an encoder,
which maps the message $m\in\{1,2,\dots,M\}$ to a \emph{desired electron intensity} $\lambda(t)$, taking binary values $\lambda(t)\in \{\lambda_{\min},\lambda_{\max}\},t\in[0,T]$,\footnote{Herein, we restrict the input signal to take binary values, since this choice is optimal \cite{MicheIT}.}
where
$T$ is the codeword duration,
and $\lambda_{\min}>0$ and $\lambda_{\max}>\lambda_{\min}$ are, respectively, the minimum and maximum electron intensities allowed into the cable.

The channel state $E(t)$ evolves in a stochastic fashion as a result of 
electrons randomly entering and exiting the cable.
Electrons enter the bacterial cable following a Poisson process with rate $\lambda_{in}(t){=}\alpha(E(t))\lambda(t)$, resulting in 
the state $E(t)$ to increase by one unit,
where $\alpha(i)\in[0,1]$ is the clogging function
and $\lambda(t)$ is the desired electron intensity, to be designed.
Electrons exit the cable  following a Poisson process with rate $\mu(E(t))$, resulting in 
the state $E(t)$ to decrease by one unit.
In general, $\alpha(i)$ is a non-increasing function of $i$,
with $\alpha(E_{\max}){=}0$, $\alpha(0){=}1$, $\alpha(i){>}0,\forall i{<}E_{\max}$.
In fact, the larger the amount of electrons carried by the cable $E(t)$,
 the more severe ATP saturation and electron clogging, and thus the smaller $\alpha(E(t))$.
 
 \subsubsection{Capacity analysis}
 \label{capacityanalysis}
 
 Using results on the capacity of finite-state Markov channels~\cite{Chen},
we have proved that the capacity of bacterial cables is given by
\begin{align}
\label{cap3}
\mathcal I
=
\max_{\bar\lambda:\mathcal E \mapsto [\lambda_{\min},\lambda_{\max}]}
\sum_{i=0}^{E_{\max}}\pi_{\bar\lambda}(i) I(\bar\lambda(i);i),
\end{align}
where
\begin{itemize}
\item $\bar\lambda(i)$ is the average desired input electron intensity in state $E(t)=i$,
defined as
\begin{align}
\label{barlambda}
\bar\lambda(i)
\triangleq
\mathbb E\left[\lambda(t)|E(t)=i\right];
\end{align}
\item
$I(\bar\lambda(i);i)$ is the \emph{instantaneous mutual information rate} in state $E(t)=i$ with expected 
desired input electron intensity $\bar\lambda(i)$,
given by
\begin{align}
\label{imu}
I(x;i)
\triangleq
&
\alpha(i)x
\log_2\left(\frac{\lambda_{\max}}{x}\right)
\nonumber\\&
+\alpha(i)\frac{\lambda_{\max}-x}{\lambda_{\max}-\lambda_{\min}}
\lambda_{\min}
\log_2\left(\frac{\lambda_{\min}}{\lambda_{\max}}\right);
\end{align}
\item
$\pi_{\bar\lambda}(i)$ is the asymptotic steady-state distribution of the bacterial cable, induced by the
desired input electron intensity $\bar\lambda(\cdot)$, given by
\begin{align}
\label{SSD}
\pi_{\bar\lambda}(i)
=
\prod_{k=0}^{i-1}\frac{\alpha(k)\bar\lambda(k)}{\mu(k+1)}
\pi_{\bar\lambda}(0),\!\!
\end{align}
and $\pi_{\bar\lambda}(0)$ is obtained via normalization ($\sum_i\pi_{\bar\lambda}(i)=1$).
\end{itemize}

The capacity optimization problem in Eq.~(\ref{cap3}) is a Markov decision process \cite{Bertsekas2005},
with state $E(t)$, action $\bar\lambda(i)$ in state $E(t)=i$, which generates the binary intensity signal,
 and reward function $I(\bar\lambda(i);i)$ in state $E(t)=i$,
and can thus be solved efficiently using standard optimization algorithms, \emph{e.g.}, policy iteration (see \cite{Bertsekas2005}).
The following trade-off arises:
1) the optimal input signal should, on the one hand,
achieve high instantaneous information rate, \emph{i.e.}, it should maximize $I(\bar\lambda(E(t));E(t))$ at each time instant $t$;
2) on the other hand, it should
induce an "optimal" steady-state distribution of the cable, 
such that those states characterized by less severe clogging and
where the transmission of information is more favorable are visited more frequently.
These two goals are in tension. In fact, the instantaneous information rate is maximum in states with large clogging state $\alpha(i)\simeq 1$, \emph{i.e.},
when $E(t)$ is small and the bacterial cable is deprived of electrons.
Visits to these states are achieved more frequently by choosing $\lambda(t)=\lambda_{\min}$ with probability one.
However, under a deterministic input distribution the  information rate is zero.

 \begin{figure}
\begin{center}
\includegraphics*[width=0.9\linewidth]{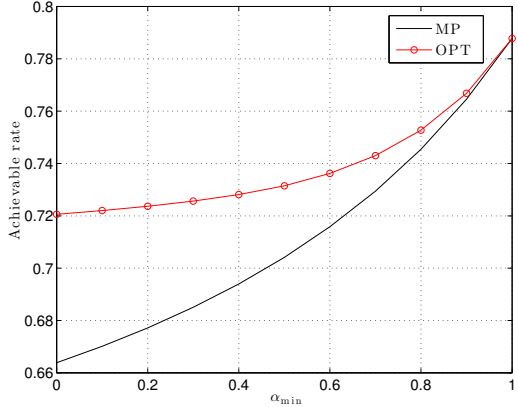}
\vspace{-3mm}
\caption{Achievable information rate as a function of the minimum clogging state value, $\alpha_{\min}$, for the
MP and OPT input distributions \cite{MicheIT}.}
\label{figXIT}
\end{center}
\vspace{-7mm}
\end{figure}

\subsubsection{Myopic signaling}
\label{myopicsignaling}
While the optimal signaling $\bar\lambda^*(i)$ balances this tension by
giving up part of the instantaneous information rate in favor of better states in the future,
the \emph{myopic} signal greedily maximizes the instantaneous information rate, without considering 
its impact on the steady-state distribution of the cable.
This is defined, $\forall i\in\mathcal E$, as
\begin{align}
\label{eqmp}
\!\!\bar\lambda_{MP}(i)\triangleq\!\!\!\!\underset{x\in[\lambda_{\min},\lambda_{\max}]}{\arg\max}\!\!\!I(x;i)
{=}\frac{\lambda_{\max}}{e}\left(\frac{\lambda_{\min}}{\lambda_{\max}}\right)^{
-\frac{\lambda_{\min}}{\lambda_{\max}-\lambda_{\min}}},\!
\end{align}
and is constant with respect to the cable state $E(t)=i$,
so that it does not require channel state information at the encoder.
Importantly, the myopic signal neglects the steady-state behavior of the cable,
and thus it tends to quickly recharge the ATP reserves of the cells,
resulting in severe clogging of the cable.
Thus, the myopic signaling may induce frequent visits to states characterized by small clogging state
 $\alpha(E(t))\simeq 0$, \emph{i.e.},  when $E(t)$ approaches $E_{\max}$.
 
 Indeed, in \cite{MicheIT} we have shown that 
  any input distribution $\bar\lambda(i)$ larger than the myopic one $\bar\lambda_{MP}(i)$
is deleterious to the capacity for the following two reasons:
1) a lower instantaneous mutual information rate is achieved, compared to $\bar\lambda_{MP}(i)$ (by definition of the myopic distribution, which maximizes $I(x;i)$);
2) faster recharges of electrons within the cable are induced,
resulting in frequent clogging of the cable, where the instantaneous information rate is small.

\subsubsection{Numerical results}

We consider a cable with electron capacity $E_{\max}=1000$.
The clogging state $\alpha(i)$ and output rate $\mu(i)$ are given by\footnote{The specific choices of $\alpha(i)$ and $\mu(i)$ have been discussed with 
Prof. M. Y. El-Naggar and S. Pirbadian, Department of Physics and Astronomy, University of Southern California, Los Angeles, USA.}
\begin{align}
\label{A}
&\alpha(i)=
\chi(i<E_{\max})\left[
1-(1-\alpha_{\min})\frac{i}{E_{\max}}\right],
\\&\nonumber
\mu(i)=0.6+0.8\frac{i}{E_{\max}}.
\end{align}

In Fig.~\ref{figXIT}, we note that the achievable information rate increases with $\alpha_{\min}$ 
under both the optimal signaling (OPT) and the myopic signaling (MP).
This is because, as $\alpha_{\min}$ increases, both $\alpha(i)$ and the instantaneous information rate $I(x;i)$ increase as well (see Eqs. (\ref{A}) and (\ref{imu})).
Intuitively,  the larger $\alpha(i)$, the better the ability of the  bacterial cable to transport electrons.

OPT outperforms MP by $\sim 9\%$ for small values of $\alpha_{\min}$. In fact, clogging is severe when the cable state approaches the maximum value $E_{\max}$.
 Therefore, in order to achieve high instantaneous information rate, the
state of the cable $E(t)$ should be kept small.
MP greedily maximizes the instantaneous information rate, but this action results in an unfavorable steady-state distribution, 
such that the cable is often in large queue states $E(t)\simeq E_{\max}$, where clogging is severe and most electrons are dropped at the cable input.
On the other hand, OPT gives up some instantaneous information rate  in order to favor the occupancy of 
low queue states, where $\alpha(i)$ approaches one and the transfer of information is maximum.
The performance degradation of MP with respect to OPT is less severe when $\alpha_{\min}{\to}1$. In fact, in this case
the instantaneous information rate $I(x;i)$ is the same in all states (except $E_{\max}$, where $\alpha(E_{\max}){=}0$ and $I(x;E_{\max}){=}0$),
hence optimizing the steady-state distribution of the cable is unimportant.
Further discussion can be found in \cite{MicheIT}.
\vspace{-3mm}
\section{Towards a Model of Quorum Sensing in a homogeneous cell colony}
\label{QSmodel}
\begin{figure}
\begin{center}
\includegraphics*[width=0.9\linewidth]{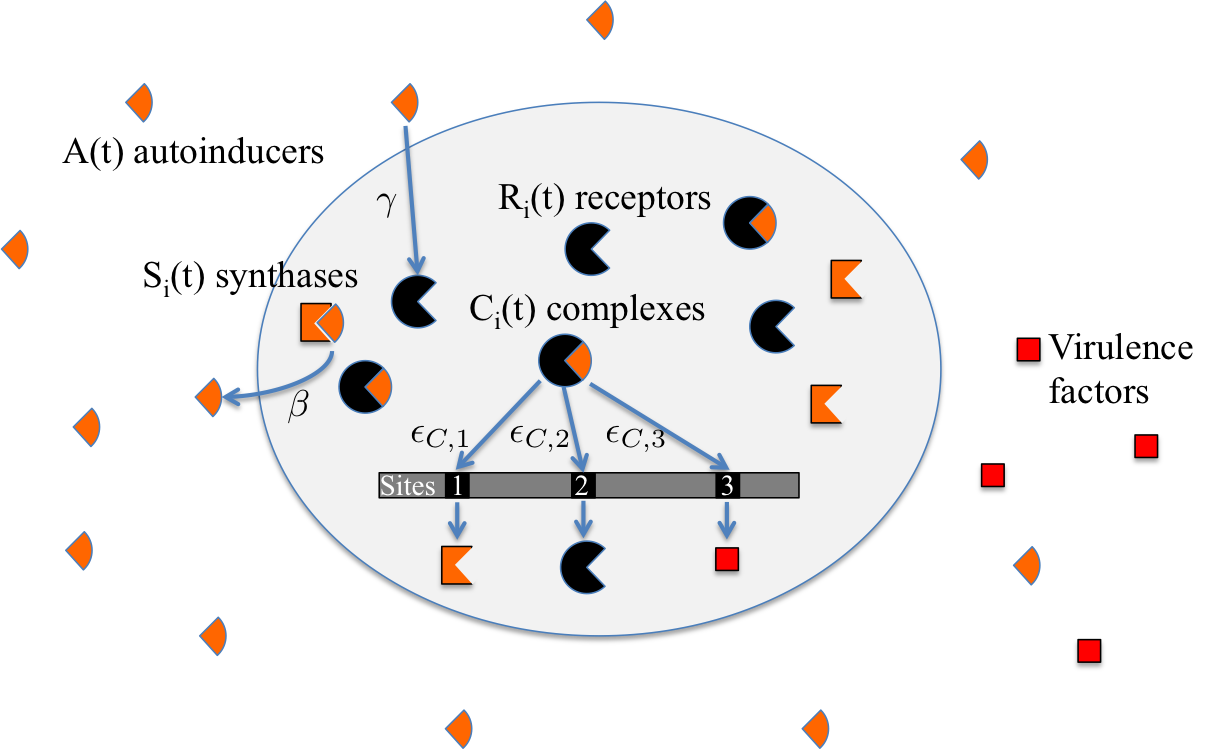}
\caption{Quorum sensing signaling system.}
\label{figQSmodel_a}
\end{center}
\vspace{-7mm}
\end{figure}

\begin{figure}
\begin{center}
\includegraphics*[width=0.9\linewidth]{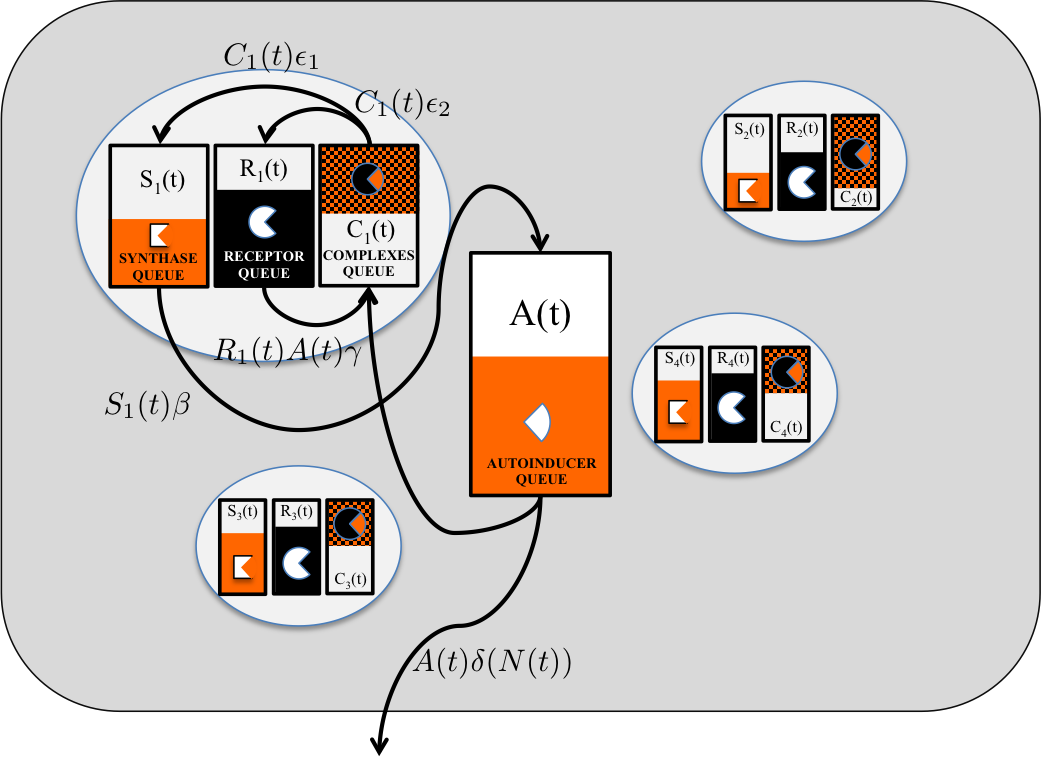}
\vspace{-3mm}
\caption{Queuing model of quorum sensing signaling system.}
\label{figQSmodel_b}
\end{center}
\vspace{-7mm}
\end{figure}

In this paper, we propose stochastic queuing models as a general framework to capture signal interactions in microbial communities.
We apply this general framework to two different microbial systems: \emph{electron transfer over bacterial cables} and \emph{quorum sensing}.
In the previous section, we summarized a queuing model of electron transfer over bacterial cables \cite{JSACmiche}, and we have shown how 
complexity reduction can be achieved by a more compact queuing model, which represents the number of electrons carried in the cable by the state variable $E(t)$
 and defines a \emph{clogging function} $\alpha(E(t))$ over its state space \cite{MicheIT}. 
Similarly, in this section, we propose a queuing model for the simplest quorum sensing signaling, for the case of a homogeneous cell colony which uses
a single autoinducer-receptor pair, depicted in Figs.~\ref{figQSmodel_a} and~\ref{figQSmodel_b}.
This is the case, for instance, for \emph{Chromobacterium violaceum}~\cite{Stauff}.
Similarly to the queuing model of electron transfer,
this model can then be used as a baseline for further model reduction techniques.

We consider a colony composed of $N(t)$ identical cells at time $t$. 
We assume that these cells can duplicate, but do not die.
In fact, only a small fraction of cells under favorable growth conditions are dead ($<0.01\%$) \cite{Stewart}, therefore cell death has a negligible influence on quorum sensing dynamics.
We model cell duplication as a Poisson process with intensity $\rho(n)$ per cell,
function of the cell population size $N(t)=n$,
 \emph{i.e.}, each cell duplicates once every $1/\rho(n)$ hours, on average.
Therefore, $N(t)$ is a counting process with time-varying intensity $N(t)\rho(N(t))$.
One model of $\rho(n)$ is the \emph{logistic growth model} $\rho(n)=\rho_{\max}\left(1-n/N_{\max}\right)$ \cite{Dilanji,Pai},
where $\rho_{\max}$ is the maximum growth rate and $N_{\max}$ is the maximum population size.
This model presumes that growth slows down as the population increases, until $N(t)=N_{\max}$ when growth stops.
We denote the volume of a single cell as $\phi_{cell}$, so that the total cell volume 
of a colony of $N(t)=n$ cells is $V_{cell}(n)=n\phi_{cell}$,
 and the total volume occupied by the colony (cell volume and extracellular environment) as $V_{tot}(n)$.
Here we consider both closed systems, with a fixed total volume $V_{tot}(n)=\text{constant}$, and open systems,
in which the volume changes with the number of cells,
$V_{tot}(n)=n\phi_{ex}$,
where $\phi_{ex}$ is the volume occupied by each cell (cell volume and extracellular environment).

The quorum sensing signaling system is characterized by different signals, represented in Fig.~\ref{figQSmodel_a}:
\emph{autoinducer molecules}, produced by the \emph{synthases} within each cell and released in the environment;
\emph{receptors}, located within each cell, which bind with the autoinducer molecules to form \emph{autoinducer-receptor complexes};
these complexes then bind to the active DNA sites to produce further synthases, receptors and virulence factors.
The dynamics of each of these signals and their interaction are characterized herein.
\vspace{-3mm}
\subsection{Environment state}
\label{envstate}
Let $A(t)$ be the number of autoinducer molecules in the system.
Thus, the concentration of autoinducers in the volume occupied by the colony is
\begin{align}
\label{autoindconcent}
\eta_A(t)\triangleq \frac{A(t)}{V_{tot}(N(t))}.
\end{align}
The dynamics of $A(t)$ over time is 
affected by autoinducer production, degradation, receptor binding, and leakage.
Leakage accounts for the autoinducer molecules that leak out of the volume before being captured by the cells.
We assume that the larger the colony, the smaller the leakage. In fact, the larger the colony of cells, the higher the likelihood  that these molecules are detected
before they diffuse out of the system.
Additionally, autoinducer molecules may chemically degrade thus becoming inactive.
We thus let $\delta_A(n)$ be the rate of leakage and degradation for each autoinducer molecule as a function of the number of cells $N(t){=}n$, where $\delta_A(n){\geq}\delta_A(n+1),\forall n{\geq}1$. An example is 
\begin{align}
\label{deltaAn}
\delta_A(n)=\xi_D+\frac{\xi_{L,1}}{1+\xi_{L,2}(n-1)},
\end{align}
where $\xi_D$ is the degradation rate of each autoinducer molecule,
 $\xi_{L,1}$ is the leakage rate when only one cell is present,
and $\xi_{L,2}$ is the rate of leakage decay as the colony size increases.
Given that there are $A(t)$ autoinducer molecules in the system, the overall leakage rate is $A(t)\delta_A(N(t))$.
 \begin{remark}
  Both $\xi_{L,1}$ and $\xi_{L,2}$
 depend on the diffusion properties of the medium and the spatial distribution of the cells \cite{Kastrup}.
  For a closed system without leakage, we have that $\xi_{L,1}=0$, hence $\delta_A(n)=\xi_D,\forall n$.
 However, (\ref{deltaAn}) may not capture all possible scenarios.
 Other scenarios can be accommodated by an appropriate selection of $\delta_A(n)$ (possibly, different from (\ref{deltaAn})).
 \end{remark}
 \vspace{-3mm}
 \subsection{Cell state}
Each cell is described by three state variables, as represented in Fig.~\ref{figQSmodel_b}. For cell $i$, we let $R_i(t)$ be the number of unbound receptors at time $t$,
and $C_i(t)$ be the number of autoinducer-receptor complexes within the cell.
Each cell uses a protein called
\emph{synthase} to synthesize autoinducers. In turn, these autoinducers are released outside the cell through its membrane.
We let $S_i(t)$ be the number of synthases within cell $i$ at time $t$.
The state of cell $i$ is thus given by $(R_i(t),C_i(t),S_i(t))$ at time $t$.
No leakage of receptors, complexes or synthases outside of the cell occurs.

Each of the $C_i(t)$ autoinducer-receptor complexes randomly binds to active  sites in the DNA sequence, resulting in specific gene expression (see Fig.~\ref{figQSmodel_a}).
We assume three active binding sites: 
 1) the first site contains the code to produce
 \emph{synthases}, resulting in the increase of $S_i(t)$ by one unit;
2) the second site contains the code to produce receptors, resulting in the increase of $R_i(t)$ by one unit;
3) the third site contains the code to produce the virulence factor.
\begin{remark}
It is worth noting that, while in the case of \emph{Chromobacterium violaceum}
 the receptor and the synthase are made from distinct DNA binding sites, in some cases
 both compounds are made from the same DNA binding site \cite{Gray}.
 This scenario can be incorporated by letting
  the first site contain the code to synthesize \emph{both} synthases and  receptors,
  resulting in the increase of \emph{both} $S_i(t)$ and $R_i(t)$ by one unit.
  Additionally, individual binding events may lead to the production of multiple synthases or receptors \cite{Teng2}.
   This scenario can be included by defining a random number of synthases or receptors produced at each binding event, with a given
   probability distribution.
 \end{remark}
 
 After the gene is expressed, the autoinducer-receptor complex unbinds from the DNA site. We assume that the binding-unbinding process is instantaneous,
\emph{i.e.}, we neglect the amount of time that the complex occupies the DNA site.
We model the process of autoinducer-receptor complexes binding to the DNA sites as Poisson processes,
whose intensities depend on the affinity of the complex to a specific site.
Thus, we let $\epsilon_{C,j},j\in\{1,2,3\}$ be the binding rate of each complex to site $j$ (see Fig.~\ref{figQSmodel_a}).
Additionally, gene expression occurs even in the absence of complexes, at basal rate 
$\epsilon_{0,j}, j\in\{1,2,3\}$.
Therefore,
since there are $C_i(t)$ autoinducer-receptor complexes within cell $i$,
 the gene located in site $j$ is expressed with overall rate 
$\epsilon_{0,j}+C_i(t)\epsilon_{C,j}$.
\vspace{-6mm}
\subsection{State dynamics}
$N(t)$ \emph{increases} over time due to cell growth;
$A(t)$ \emph{increases} over time as the synthases located within each cell produce autoinducers,
and \emph{decreases} over time as a result of autoinducers binding to receptors, leakage and degradation (Sec.~\ref{envstate});
$R_i(t)$ \emph{increases} over time as the cell synthesizes receptors,
and \emph{decreases} over time as a result of autoinducers binding to receptors and receptor degradation;
similarly, 
$C_i(t)$ \emph{increases} over time as the unbound receptors bind to the autoinducer signal in the environment,
and decreases over time as a result of degradation and complex unbinding;
finally, $S_i(t)$ increases over time as the cell produces synthases,
and decreases over time as a result of degradation.
We let 
\begin{align}
\left\{\begin{array}{l}
R_{\text{TOT}}(t)\triangleq\sum_{i=1}^{N(t)}R_i(t),\\
C_{\text{TOT}}(t)\triangleq\sum_{i=1}^{N(t)}C_i(t),\\
S_{\text{TOT}}(t)\triangleq\sum_{i=1}^{N(t)}S_i(t),
\end{array}\right.
\end{align}
be the total amount of  receptors, complexes and synthases of the cell colony, 
and
\begin{align}
\label{concentrationofothers}
\left\{\begin{array}{l}
\eta_R(t)\triangleq\frac{R_{\text{TOT}}(t)}{V_{cell}(N(t))},\\
\eta_C(t)\triangleq\frac{C_{\text{TOT}}(t)}{V_{cell}(N(t))},\\
\eta_S(t)\triangleq\frac{S_{\text{TOT}}(t)}{V_{cell}(N(t))},
\end{array}\right.
\end{align}
be their concentrations, respectively.\footnote{Note that receptors, complexes and synthases are located inside the cell rather then in the extracellular environment,
therefore their concentrations are calculated with respect to the total cell volume $V_{cell}(N(t))$.}
There are several processes that affect the dynamics of $(N(t),A(t),R_{\text{TOT}}(t),C_{\text{TOT}}(t),S_{\text{TOT}}(t))$, as detailed below.

\paragraph{Cell duplication}
Cell duplication occurs with rate $\rho(N(t))$ for each cell, so that the cell population $N(t)$ increases by one unit with rate $\rho(N(t)) N(t)$.
 When one cell duplicates, its content is split randomly among the two cells.
Therefore, if $(R_i(t),C_i(t),S_i(t))$ is the state of cell $i$ before its duplication,
after the duplication (at time instant $t^+$) there are two cells with state $(R_i^{(1)}(t^+),C_i^{(1)}(t^+),S_i^{(1)}(t^+))$ and $(R_i^{(2)}(t^+),C_i^{(2)}(t^+),S_i^{(2)}(t^+))$, respectively,
where $R_i^{(1)}(t^+)+R_i^{(2)}(t^+)=R_i(t)$, $C_i^{(1)}(t^+)+C_i^{(2)}(t^+)=C_i(t)$, 
and  $S_i^{(1)}(t^+)+S_i^{(2)}(t^+)=S_i(t)$,
 with probability distribution
\begin{align}
&\!\!\mathbb P\!\left(\!(R_i^{(1)}(t^+),C_i^{(1)}(t^+),S_i^{(1)}(t^+)){=}(r,c,s)|R_i(t),C_i(t),S_i(t)\!\right)
\nonumber\\&
\!\!=
\left(\!\begin{array}{c}R_i(t)\\r\end{array}\!\right)
\left(\!\begin{array}{c}C_i(t)\\c\end{array}\!\right)
\left(\!\begin{array}{c}S_i(t)\\s\end{array}\!\right)
2^{-R_i(t)-C_i(t)-S_i(t)}.
\end{align}

 \begin{table*}[t]
\caption{Description of the queuing abstraction for the quorum sensing model.}
\vspace{-3mm}
\label{tableQS}
\begin{center}
\footnotesize
\scalebox{1}{
\begin{tabular}{| c | c | c | l |}
 \hline	
\bf Queue & \bf Arrival rate & \bf Service rate & \bf Explanation\\ \hline
$N(t)$ & $N(t)\rho(N(t))$ & $0$ & cell population size\\ \hline
$A(t)$ & $\beta S_{\text{TOT}}(t)+\upsilon_CC_{\text{TOT}}(t)$
&  $\delta_A(N(t))A(t)+\lambda_C(A(t),R_{\text{TOT}}(t),N(t))$ & number of autoinducer molecules \\ \hline
$R_{\text{TOT}}(t)$ &
$N(t)\epsilon_{0,2}+C_{\text{TOT}}(t)\epsilon_{C,2}+\upsilon_CC_{\text{TOT}}(t)$
& $\delta_RR_{\text{TOT}}(t)+\lambda_C(A(t),R_{\text{TOT}}(t),N(t))$
  & number of receptors  \\ \hline
$C_{\text{TOT}}(t)$  & $\lambda_C(A(t),R_{\text{TOT}}(t),N(t))$
& $\delta_CC_{\text{TOT}}(t)+\upsilon_CC_{\text{TOT}}(t)$
& number of complexes \\ \hline
$S_{\text{TOT}}(t)$  & $N(t)\epsilon_{0,1}+C_{\text{TOT}}(t)\epsilon_{C,1}$ & $\delta_SS_{\text{TOT}}(t)$ &
 number of synthases \\ \hline
\end{tabular}
}
\end{center}
\vspace{-7mm}
\end{table*}

\paragraph{One autoinducer is created}
Each synthase produces autoinducers with rate $\beta$.
Since there are $S_{\text{TOT}}(t)$ synthases,
the cell colony produces autoinducers with rate $\beta S_{\text{TOT}}(t)$,
resulting in the increase of $A(t)$ by one unit.

\paragraph{One autoinducer leaks or degrades}
Each autoinducer leaks or degrades with rate $\delta_A(N(t))$.
Therefore, autoinducer leakage and degradation occurs with rate
$\delta_A(N(t))A(t)$,  resulting in the
decrease of $A(t)$ by one unit. Degradation rates are dependent on auto-inducer type and environmental factors, as discussed in \cite{Pai,Surette}.

\paragraph{One receptor is created}
One receptor is created whenever one complex binds to the second active DNA site.
This occurs with rate $N(t)\epsilon_{0,2}+C_{\text{TOT}}(t)\epsilon_{C,2}$ across the whole cell colony,
resulting in the increase of $R_{\text{TOT}}(t)$ by one unit.

\paragraph{One receptor degrades}
Each receptor degrades with rate $\delta_R$.
Therefore, receptor degradation occurs with rate
$\delta_RR_{\text{TOT}}(t)$,  resulting in the decrease of $R_{\text{TOT}}(t)$ by one unit.

\paragraph{One complex is created}
Typically, binding of autoinducers to receptors to form autoinducer-receptor complexes does not occur until a threshold concentration of autoinducers is achieved \cite{PerezVelazquez}.
We let this threshold be $\eta_{A,\text{th}}$. Thus, if $\eta_A(t)<\eta_{A,\text{th}}$, then 
no binding of autoinducers to receptors occurs. When the threshold is exceeded, \emph{i.e.}, $\eta_A(t)\geq \eta_{A,\text{th}}$, then
the binding of autoinducers to receptors is a Poisson process
with intensity $\gamma$ [per unit of autoinducer and receptor concentrations, per unit volume, per hour]. Since the concentrations
of autoinducers and free receptors is $\eta_A(t)$ and $\eta_R(t)$, respectively, 
and the total cellular volume where these reactions occur is $V_{cell}(N(t))=N(t)\phi_{cell}$,
complexes form with rate
\begin{align}
\label{ratecomplexes}
&\lambda_C(A(t),R_{\text{TOT}}(t),N(t))
\nonumber\\&
\triangleq
\gamma \eta_A(t)\eta_R(t)\chi(\eta_A(t)\geq \eta_{A,\text{th}})V_{cell}(N(t))
\nonumber\\&
= 
\gamma \frac{A(t)R_{\text{TOT}}(t)}{V_{tot}(N(t))}\chi(A(t)\geq \eta_{A,\text{th}}V_{tot}(N(t))),
\end{align}
so that  $A(t)$ and $R_{\text{TOT}}(t)$ decrease by one unit,\footnote{In this paper, for simplicity, we assume that
one receptor binds to one autoinducer to form one complex. However, in most systems, two receptors bind an autoinducer.}
and $C_{\text{TOT}}(t)$ increases by one unit.

\paragraph{One complex degrades}
Each autoinducer-receptor complex  degrades with rate $\delta_C$.
Therefore, autoinducer-receptor complex degradation occurs with rate
$\delta_CC_{\text{TOT}}(t)$,  resulting in the decrease of $C_{\text{TOT}}(t)$ by one unit.

\paragraph{One complex unbinds}
Each complex may randomly unbind, and the constituent autoinducer and receptor molecules become active again.
This event occurs with rate $\upsilon_C$ for each complex.
Therefore, unbinding of complexes occurs with rate $\upsilon_CC_{\text{TOT}}(t)$,  resulting in the decrease of $C_{\text{TOT}}(t)$ by one unit
and in the increase of $A(t)$ and $R_{\text{TOT}}(t)$ by one unit.

\paragraph{One synthase is created}
One synthase is created whenever one complex binds to the first active DNA site.
This occurs with rate $N(t)\epsilon_{0,1}+C_{\text{TOT}}(t)\epsilon_{C,1}$ across the whole cell colony,
resulting in the increase of $S_{\text{TOT}}(t)$ by one unit.
Since synthases create autoinducers and autoinducer-receptor complexes create new synthases, the production of autoinducer can be considered to be autocatalytic.

\paragraph{One synthase degrades}
Each synthase degrades with rate $\delta_S$.
Therefore, synthase degradation occurs with rate
$\delta_SS_{\text{TOT}}(t)$,  resulting in the decrease of $S_{\text{TOT}}(t)$ by one unit.

\paragraph{Virulence factor expression}
Finally, the gene located in site 3 is expressed with intensity $\epsilon_{0,3}+C_i(t)\epsilon_{C,3}$ in cell $i$,
resulting in the overall expression with intensity $N(t)\epsilon_{0,3}+C_{\text{TOT}}(t)\epsilon_{C,3}$ over the whole cell colony.
\vspace{-3mm}
\subsection{State representation and state space reduction}

Let $\boldsymbol{R}(t)=[R_1(t),R_2(t),\dots,R_{N(t)}(t)]$ be the  vector of receptor concentrations,
 $\boldsymbol{C}(t)=[C_1(t),C_2(t),\dots,C_{N(t)}(t)]$ be the vector of autoinducer-receptor complexes concentrations,
 and
 $\boldsymbol{S}(t)=[S_1(t),S_2(t),\dots,S_{N(t)}(t)]$ be the vector of synthase concentrations.
These vectors have time-varying length $N(t)$, due to cell duplication.
Then, the state of the system at time $t$ is described by the tuple $(N(t),A(t),\boldsymbol{R}(t),\boldsymbol{C}(t),\boldsymbol{S}(t))$.
We notice that the state space grows exponentially with the cell population size, similarly to the model of electron transfer 
in Sec.~\ref{bactcable}.
However, the analysis of state dynamics above highlights that 
$(N(t),A(t),R_{\text{TOT}}(t),C_{\text{TOT}}(t),S_{\text{TOT}}(t))$
is a Markov chain with  dynamics described in Table \ref{tableQS}, hence complexity reduction can be achieved.
Indeed, 
this stochastic model and such Markov property can be exploited to infer the overall quorum sensing dynamics,
rather than tracking the dynamics of each specific cell, \emph{i.e.},
\begin{align}
&p_t(n,a,r,c,s)
\\&\nonumber
\triangleq
\mathbb P\left(
N(t){=}n,A(t){=}a,R_{\text{TOT}}(t){=}r,C_{\text{TOT}}(t){=}c,S_{\text{TOT}}(t){=}s
\right).
\end{align}
\section{Simulations Results and Experimental Data}
\label{simres}


In this section, we present simulation results for the homogeneous cell colony considered in Sec.~\ref{QSmodel}. 
The parameters  are listed in Table~\ref{table1}.
We consider two different setups:
\emph{open system} and \emph{closed system}, both 
with  initial state $(N(0),A(0),R_{\text{TOT}}(0),C_{\text{TOT}}(0),S_{\text{TOT}}(0))=(1,0,0,0,0)$ and the following parameters.

{\bf a) Open system:} Here, cells in an open system exist as a densely packed biofilm attached tho a surface.
Autoinducers may leak, with parameters $\xi_{L,1}=5000$ and $\xi_{L,2}=0.1$, and the cell colony grows in an open space.
Cells grow tightly packed together, so that
the extracellular environment takes only 10\% of the cell volume ($\phi_{ex}=1.1\phi_{cell}$).
The maximum population size is $N_{\max}=\infty$, so that cell growth does not slow down.
The concentration of autoinducers is computed as in (\ref{autoindconcent}), over the total volume 
$V_{tot}(N(t))=N(t)\phi_{ex}$, whereas the concentration of receptors, complexes and synthases is computed as in 
(\ref{concentrationofothers}), over the cell volume $V_{cell}(N(t))=N(t)\phi_{cell}$.
 
 \emph{\bf b) Closed system:}
Autoinducers do not leak ($\xi_{L,1}=\xi_{L,2}=0$).
The cell colony grows in a closed space of volume $V_{tot}(n)=0.1\mathrm{nL}$.
The maximum population size is $N_{\max}=1000$,
so that, when growth is complete, the cell volume takes 1\% of the total volume.
The concentration of autoinducers is computed as in (\ref{autoindconcent}), over a constant total volume 
$V_{tot}(N(t))=0.1\mathrm{nL}$, whereas the concentration of receptors, complexes and synthases is computed as in 
(\ref{concentrationofothers}), over the cell volume $V_{cell}(N(t))=N(t)\phi_{cell}$, which grows with the population size.

Open and closed systems occur both in nature and in experimental testbeds: microdroplets \cite{Boedicker,Carnes} and well-plates \cite{Boedicker2}
are examples of closed systems, whereas cells on agar plates \cite{McLean,Dilanji} or in the presence of flow \cite{Kirisits} are examples of open systems.

 \begin{table*}[t]
\caption{Simulation parameters ($\mathrm{nM}$ denotes nanomolar concentration)}
\vspace{-3mm}
\label{table1}
\begin{center}
\footnotesize
\scalebox{1}{
\begin{tabular}
{| c | l | c | c | c |} \hline	
Parameter & Explanation & Value [all of them per hour] & Reference\\ \hline
$\rho_{\max}$ & Maximum duplication rate & 1 [per cell] &  \\ \hline
$\phi_{cell}$ & Cell volume & 1 [$\mathrm{fL}$] &  \\ \hline
$\beta$ & Autoinducer generation rate & 18 [per synthase] & \cite{PerezVelazquez} \\ \hline
$\xi_D$ & Autoinducer degradation rate & 0.01 [per autoinducer] & \cite{Chandler} \\ \hline
$\eta_{A,\text{th}}$ & Autoinducer concentration threshold & 21.4 [$\mathrm{nM}$] & calculated from \cite{PerezVelazquez}\\ \hline
$\epsilon_{0,2}$ & Basal receptor generation rate & 80 [per cell] & calculated from \cite{PerezVelazquez} \\ \hline
$\epsilon_{C,2}$ & Activated receptor generation rate & 3 [per complex] & calculated from \cite{PerezVelazquez} \\ \hline
$\delta_R$ & Receptor degradation rate & 12 [per receptor] & \cite{Smith} \\ \hline
$\gamma$ & Complex generation rate & $3.5$ [per $\mathrm{nM}$ receptor \& autoinducer concentrations, per $\mathrm{fL}$] & \cite{Smith}\\ \hline
$\delta_C$ & Complex degradation rate & 1.4 [per complex] & \cite{Smith} \\ \hline
$\upsilon_C$ & Unbinding of complexes & 60 [per complex] & \cite{Smith} \\ \hline
$\epsilon_{0,1}$ & Basal synthase generation rate & 80 [per cell] & calculated from \cite{PerezVelazquez} \\ \hline
$\epsilon_{C,1}$ & Activated synthase generation rate & 3 [per complex] & calculated from \cite{PerezVelazquez} \\ \hline
$\delta_S$ & Synthase degradation rate & 1 [per synthase] & \cite{Bionumbers} \\ \hline
\end{tabular}
}
\end{center}
\vspace{-7mm}
\end{table*}

\begin{figure*}
\begin{center}
\setlength{\tabcolsep}{1mm}
\begin{tabular}{cc}
\subfigure[Cell concentration vs time.]{\includegraphics*[width=0.48\linewidth]{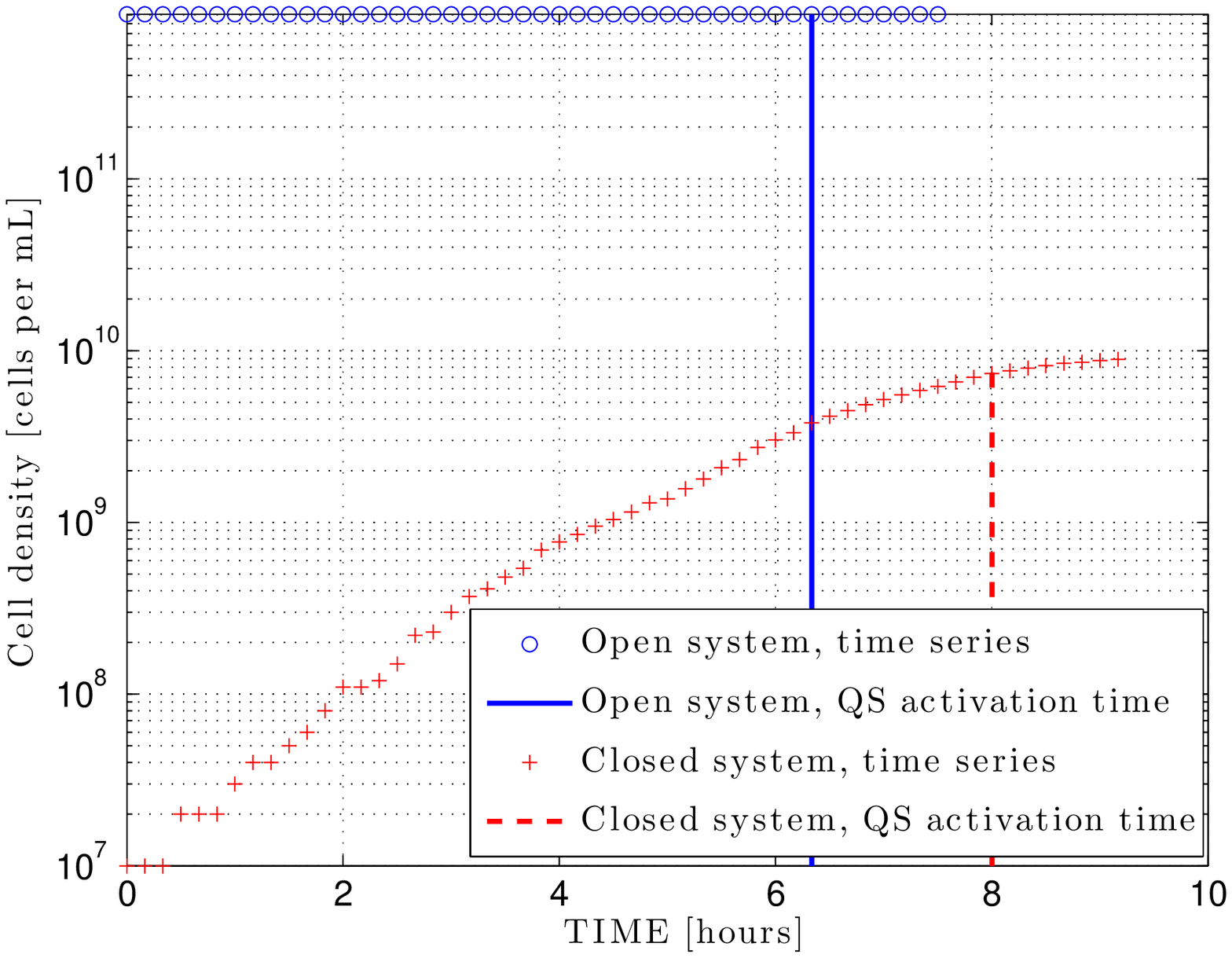}}
&
\subfigure[Autoinducers concentration vs time.]{\includegraphics*[width=0.48\linewidth]{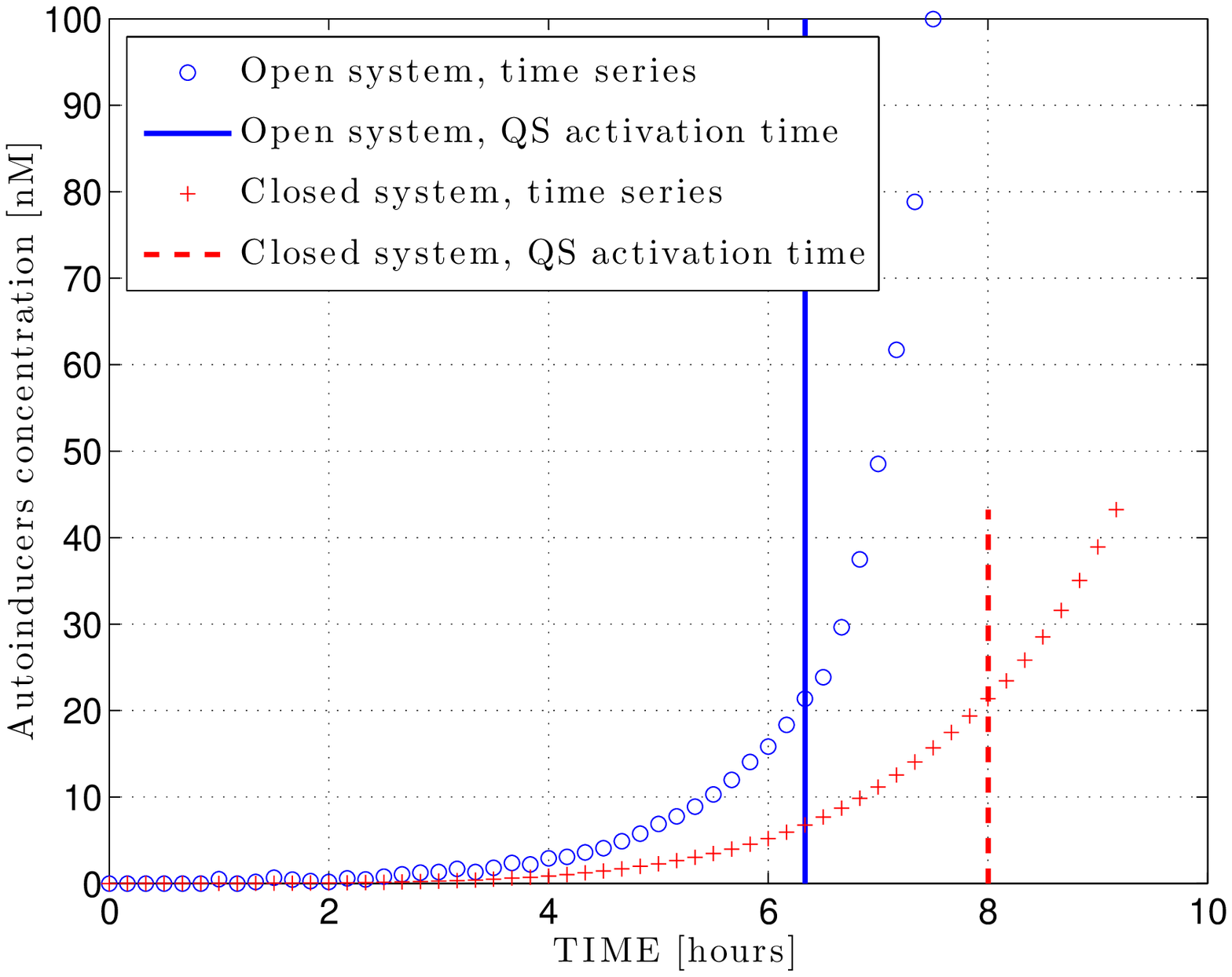}}
\\
\subfigure[Receptors concentration vs time.]{\includegraphics*[width=0.48\linewidth]{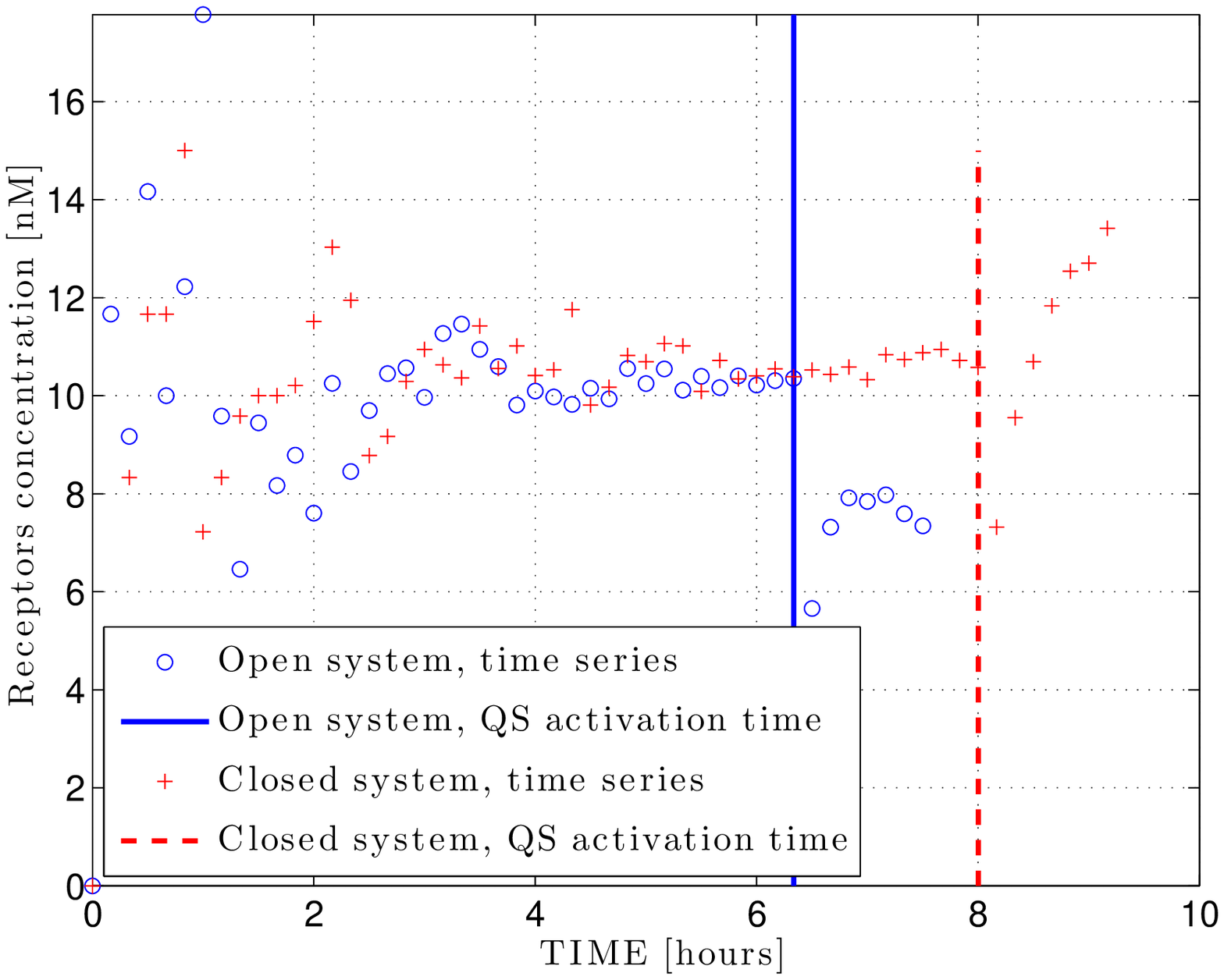}}
&
\subfigure[Autoinducer-receptor complexes concentration vs time.]{\includegraphics*[width=0.48\linewidth]{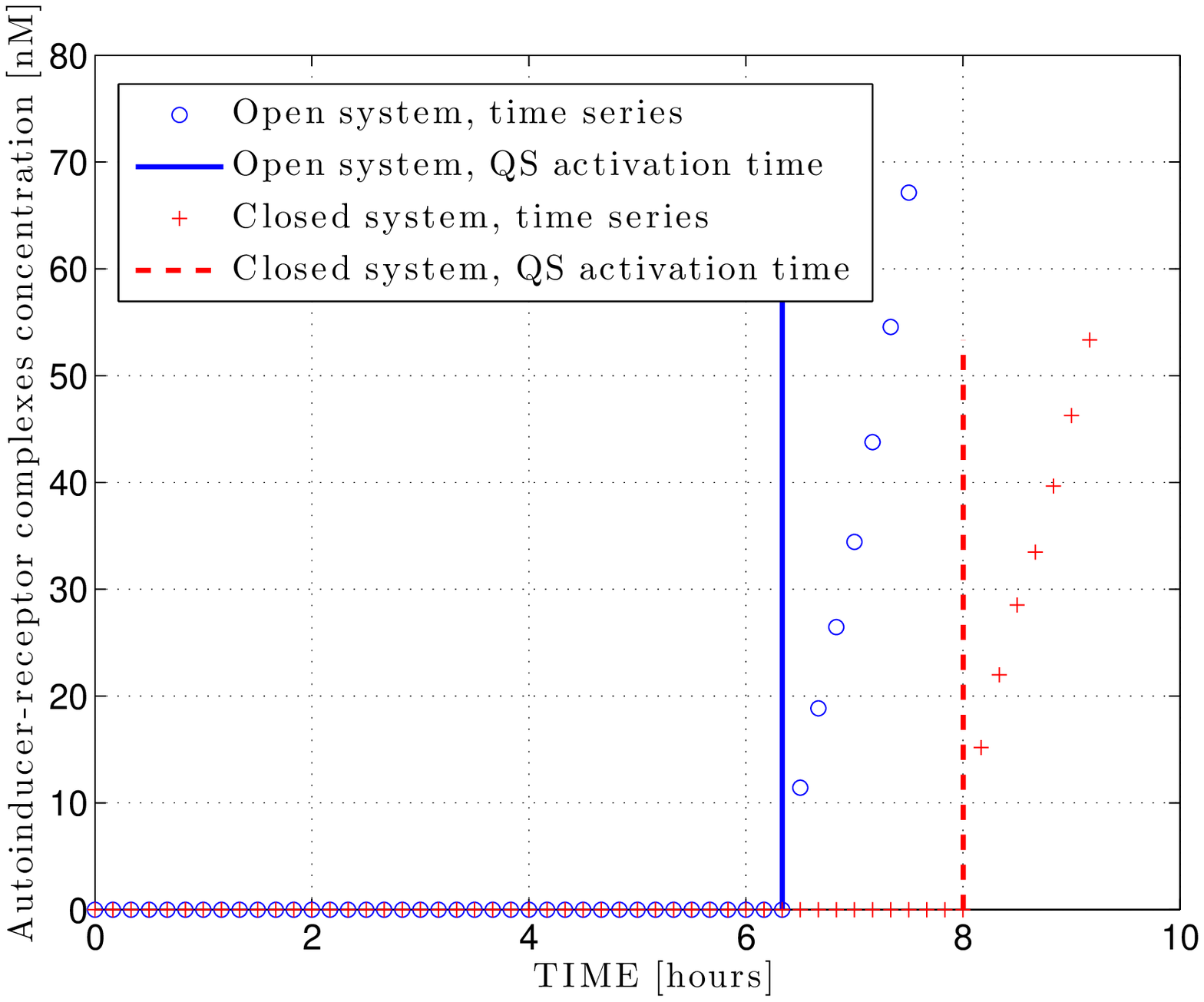}}
\\
\subfigure[Synthases concentration vs time.]{\includegraphics*[width=0.48\linewidth]{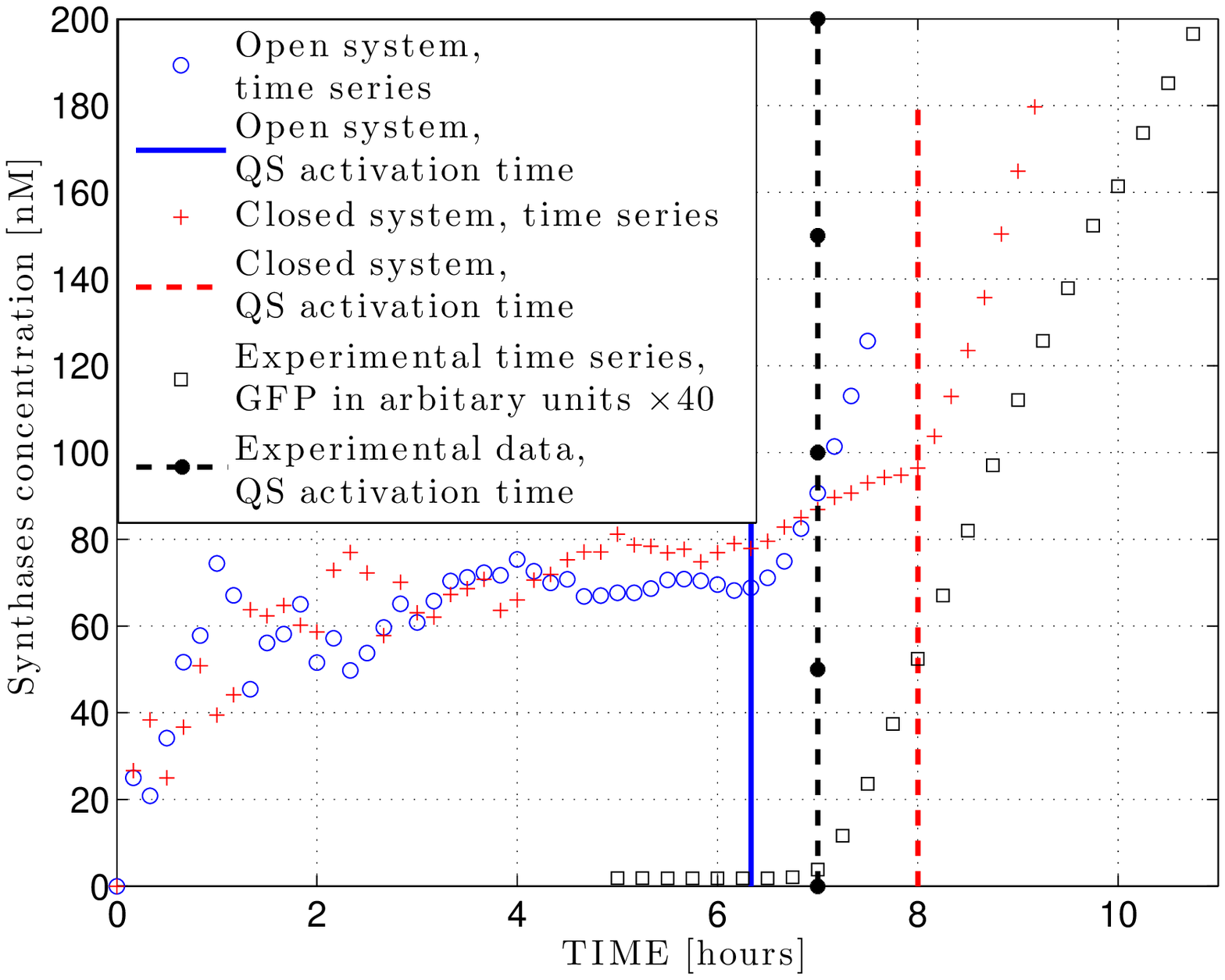}}
&
\subfigure[Experimental time-series of cell concentration, and quorum sensing activation time.]{\includegraphics*[width=0.48\linewidth]{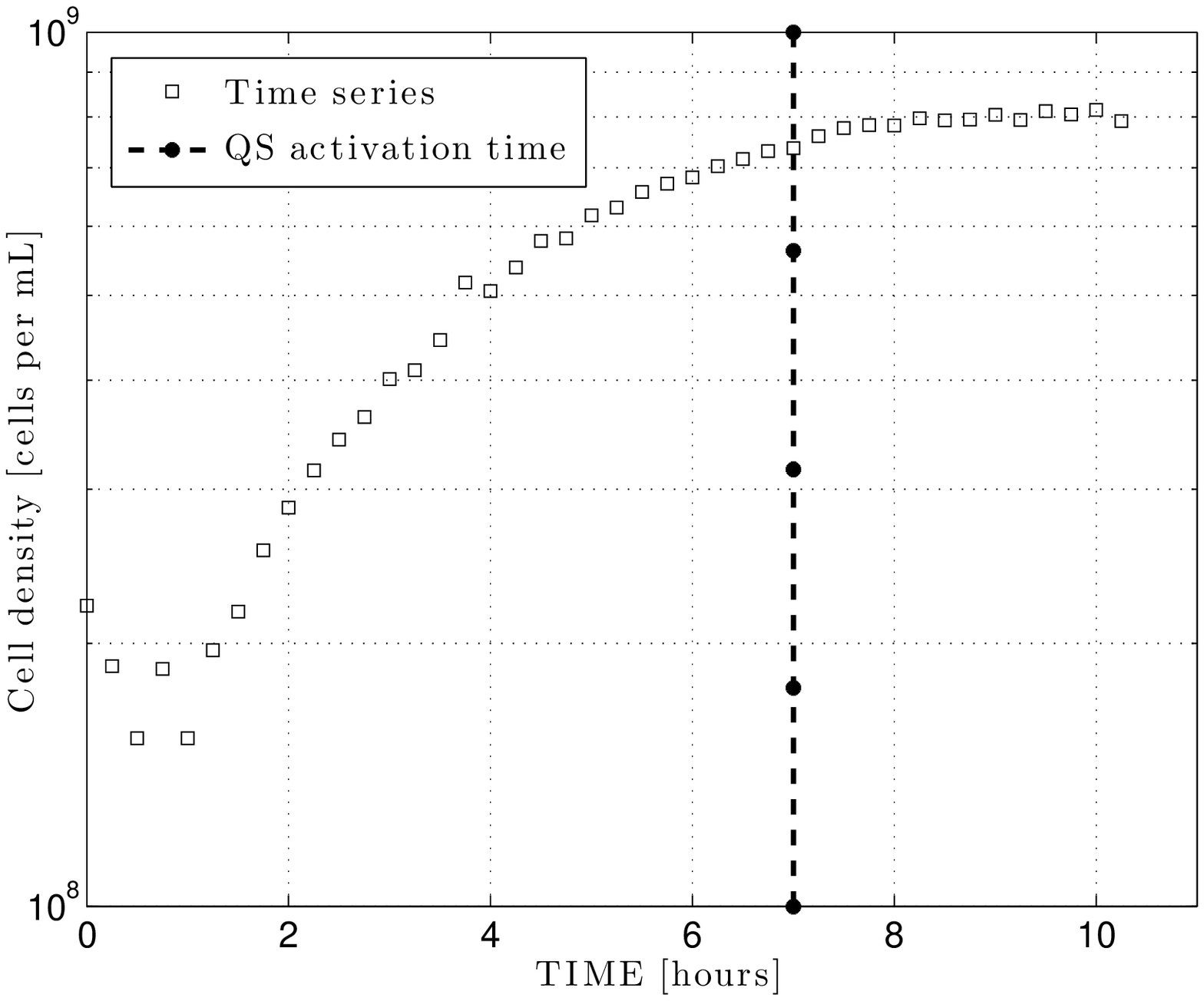}}
\end{tabular}
\caption{Simulation of quorum sensing system and experimental data. Comparison of open and closed systems. Sampling period 10m.}
\label{figQS}
\end{center}
\end{figure*}

We use Gillespie's exact stochastic simulation algorithm \cite{Gillespie_1977} to simulate the dynamics of the cell colony. In Figs.~\ref{figQS}.a-e, we plot the time series of cell concentration, the concentration of autoinducers, receptors, complexes and synthases, respectively. 
From Figs.~\ref{figQS}.d-e, we notice that the concentrations of autoinducer-receptor complexes and of synthases are small and approximately constant before 6 hours and 8 hours for the open and closed systems, respectively, and then grow quickly and steadily afterwards. We deduce that quorum sensing activation occurs after approximately 6 hours in the open system and after 8 hours in the closed system. 
Indeed, as can be observed in Fig.~\ref{figQS}.b, the concentration of autoinducers grows slowly before quorum sensing activation, 
until reaching the critical concentration $\eta_{A,\text{th}}=21.4 [\mathrm{nM}]$. Once this threshold is reached, 
complexes start to get formed with rate given by Eq. (\ref{ratecomplexes}), thus explaining the steep increase observed in Fig.~\ref{figQS}.d.
In turn, the high concentration of complexes, which frequently bind to the DNA sites, contributes to the buildup of synthases in the system,
 thus explaining the steep increase observed in Fig.~\ref{figQS}.e.
Finally, each synthase produces autoinducers with rate $\beta$, thus  contributing to the buildup of autoinducers and sustaining the quorum sensing positive feedback loop.

Quorum sensing activation is faster in the open system, since a much higher cell density is achieved (Fig. \ref{figQS}.a), resulting in a steeper growth of autoinducer concentration, and thus earlier quorum sensing activation. In the open system the parameter that accounts for the loss of autoinducers to the surroundings will depend on the system, and further work is needed to determined if closed systems in general have slower activation kinetics. 

In Fig.~\ref{figQS}.a, we note that, when quorum sensing activation occurs, cell density is much smaller in the closed system than in the open system.
This is because, in the closed system, autoinducers do not leak, and thus a lower concentration of cells is sufficient to reach the critical 
 concentration of autoinducers for activation.
In closed systems autoinducers are produced by dispersed cells and get diluted and retained in the entire volume of the system, whereas in open systems autoinducers are produced in concentrated regions of cells but escape in the surroundings. The local concentration of producers and autoinducer escape are key factors that dictate quorum sensing activation. Similar dependencies have been observed in other nonlinear systems \cite{Kastrup2}.

In Fig.~\ref{figQS}.c, we note that, in the open system, the concentration of receptors drops after quorum sensing activation. 
This is due to the high rate of autoinducers binding to receptors to form complexes, thus reducing the availability of free receptors.
A similar behavior can be noticed in the closed system. However, the drop in this case is followed by a build up of receptors. 
In fact, the lower concentration of autoinducers in this case limits the production of complexes, resulting in excess free receptors.

 
 Finally, we provide experimental data for a closed system. The genes required for quorum sensing were placed into \emph{Escherichia coli} using a plasmid constructed in \cite{Prindle}.  The genetic constructs place a green fluorescent protein under control of quorum sensing.  Cells were grown in LB media at $37\,^{\circ}{\rm C}$. 
 Growing cells are diluted such that cells remain in the early to mid
exponential phase of growth, before quorum sensing is activated, for
several generations before measurement. Since most proteins are
stable for long periods of time, maintaining cultures at low density
for several generations is needed to initialize the population of
cells to a quorum sensing off state.
  Fluorescence was monitored using a $96$ well-plate reader \cite{Boedicker2} and cell concentration was measured by absorbance of light at $600\mathrm{nm}$.
  
  As seen in Fig.~\ref{figQS}.e, the dynamics of the green fluorescent protein regulated by quorum sensing
in experiments is qualitatively similar to the predicted expression of synthase: they both exhibit the characteristic switch from a low to high production rate after quorum sensing is activated. However, green fluorescent protein production is immeasurable or zero prior to quorum sensing activation, which suggests that it has a lower basal level of gene expression than that assumed for synthase in our model. The basal level of expression can be independently tuned for each quorum sensing regulated gene. Genes that are part of the quorum sensing machinery, such as synthases and receptors, likely need a higher basal expression rate to prime quorum sensing activation, The consequences of differences in the basal and activated rates of expression for quorum sensing controlled genes warrants further investigation.

Fig.~\ref{figQS}.f plots the cell concentration over time and quorum sensing activation time. Note that the dynamics of cell concentration over time for the closed system (Fig.~\ref{figQS}.a) are similar to experimental data (Fig.~\ref{figQS}.f), \emph{i.e.}, the cell population grows exponentially fast until reaching a certain maximum concentration, thus validating the logistic growth model employed. However, both the cell duplication rate and the maximum concentration exhibit different values in our model ($\rho_{\max}=1$ [per cell per hour] and $N_{\max}/V_{tot}(n)=10^{10}$ [cells per mL], respectively) and in the experimental data ($\rho_{\max}=0.65$
[per cell per hour] and $N_{\max}/V_{tot}(n)=8\times 10^{8}$ [cells per mL], respectively). 
Indeed, these parameters are variable and depend on specifics of each experiment, such as the cell
type, composition of the media, and growth conditions.

\vspace{-3mm}
\section{Extensions}
\label{extensions}
Our  framework can accommodate several aspects of quorum sensing observed in nature.
\vspace{-3mm}
\subsection{Signal Integration}
Indeed,
most quorum sensing signaling systems are more complex than the scenario considered in Sec.~\ref{QSmodel}, which includes one receptor-autoinducer pair only.  For instance, \emph{Pseudomonas aeruginosa} contains multiple quorum sensing networks that work together to monitor changes in the local population of cells.  Since the initial discovery of the LasI/LasR system, two other quorum sensing circuits in \emph{Pseudomonas aeruginosa} have been discovered and characterized, the RhlI/RhlR and PQS \cite{Jimenez}.  Recently a fourth quorum sensing system has been identified, IQS \cite{Lee}.  Although each of these networks has a unique signal and receptor, these four systems are coupled in a feedback network in which each signaling pair regulates the production of both the synthase and receptor of at least one other network.  \emph{Vibrio} species have similar complexity in their quorum sensing networks \cite{Teng}, integrating information from multiple signals to direct regulatory decisions.  
\vspace{-3mm}
\subsection{Interference}
 In addition to individual microbes producing multiple signals, quorum sensing systems also mediate communication between different species.  Many bacteria find themselves in diverse and crowded environments, and many of these neighboring species will also produce autoinducers.  The type of autoinducer produced by many \emph{Vibrio} and \emph{Pseudomonas} species has been identified in more than 70 different species \cite{Kimura}, and there are examples of signals or enzymes that inhibit quorum sensing activation \cite{Rampioni201460}.  Within these multispecies communities, there is significant potential for interference with the process of quorum sensing activation, \emph{i.e.}, any process which alters the ability of cells to produce, exchange, recognize, or respond to a quorum sensing autoinducer.  Several mechanisms of interference that occur naturally have been identified.
\begin{enumerate}
\item	\emph{Signal synthesis inhibition}: Although no examples of this interaction have been found in nature, a synthetic compound that prevented
acyl-homoserine lactone (HSL) production by the TofI synthase in the soil microbe \emph{Burkholderia glumae} has been identified \cite{Chan}. This compound
 occupies the active site of the synthase, preventing autoinducer production.
\item \emph{Destruction or chemical modification of the autoinducer}:  The common example of signal destruction is the AiiA enzyme isolated from \emph{Bacillus} \cite{Rampioni201460}.  
AiiA is an enzyme that degrades a common type of autoinducer, thereby
preventing the autoinducer from binding to the receptor. In another
example, the oxidoreductase from \emph{Burkholderia} is also capable of
inactivating some autoinducers through chemical modification, thereby
altering the specificity for receptor molecules.
\item \emph{Competitive binding to the receptor}:  Some species have been shown to produce small molecules that interfere with autoinducer binding to the receptor, for example a furanone compound produced by the alga \emph{Delisea pulchra} \cite{Gonzalez}.  In other cases the competition is from analogous autoinducers.  There are many versions of the autoinducer HSL, with differing carbon tails, and each bacteria produces one or more types of HSL.  For example, \emph{Pseudomonas aeruginosa} produces the autoinducers C4-HSL and 3oxo-C12-HSL, whereas \emph{Chromobacterium violaceum} produces C6-HSL \cite{Williams}.  The receptor of \emph{Chromobacterium violaceum} has been found to be activated by C6-HSL, but deactivated or inhibited by HSL with longer carbon tails, such as 3oxo-C12-HSL \cite{McClean}.  In this way, crosstalk between autoinducers produced by neighboring bacteria can influence on signal transduction and quorum sensing activation.
\end{enumerate}

Our model in Sec.~\ref{QSmodel} can be extended to include different types of interference. Consider an interfering signal with concentration $\eta_I(t)\triangleq I(t)/V_{tot}(N(t))$,
where $I(t)$ is the total amount of interfering molecules in the volume of interest.
The interfering signal is injected in the system  with rate $\mu_I$, and leaks with rate $\delta_I(N(t))$, which is a function of the colony size, similarly to autoinducers.
We may consider the following models of interference:
\begin{itemize}
\item \emph{Receptor inhibition}: the interfering signal binds with the receptors, thus competing with the autoinducer signal;
unlike the autoinducer-receptor complex, which induces specific gene expression by binding to the DNA sites,
the complex formed by the interfering signal and the receptor becomes inactive;
the binding of interfering molecules to receptors is a Poisson process
with intensity $\gamma_{IR}$ 
[per unit of interfering signal and receptor concentrations, per unit volume, per hour]. Since the concentrations
of the interfering signal and of free receptors is $\eta_I(t)$ and $\eta_R(t)$, respectively, 
the overall binding rate is $\gamma_{IR} \eta_I(t)\eta_R(t)V_{cell}(N(t))$, since these reactions occur inside the cells and the total cell volume is $V_{cell}(N(t))$; 
correspondingly, both $R_{\text{TOT}}(t)$ and $I(t)$ decrease by one unit;
\item \emph{Synthase blocking}: the interfering signal binds with the active site of the synthase, thus preventing autoinducer production;
the binding of interfering molecules to synthases is a Poisson process
with intensity $\gamma_{IS}$
[per unit of interfering signal and synthase concentrations, per unit volume, per hour].
 Since the concentrations
of the interfering signal and of synthases is $\eta_I(t)$ and $\eta_S(t)$, respectively, 
the overall binding rate is $\gamma_{IS} \eta_I(t)\eta_S(t)V_{cell}(N(t))$; 
correspondingly, both $S_{\text{TOT}}(t)$ and $I(t)$ decrease by one unit;
\item \emph{Autoinducer degradation}: the interfering signal destroys or chemically modifies the autoinducer signal, thus preventing it from binding to the receptor;
the binding of interfering molecules to autoinducers is a Poisson process
with intensity $\delta_{IA}$
[per unit of interfering signal and autoinducer concentrations, per unit volume, per hour]. Since the concentrations
of the interfering signal and of autoinducers is $\eta_I(t)$ and $\eta_A(t)$, respectively, 
and these reactions occur over the volume $V_{tot}(N(t))$ occupied by the cell colony,
the overall binding rate is $\delta_{IA} \eta_I(t)\eta_A(t)V_{tot}(N(t))$; 
correspondingly, both $A(t)$ and $I(t)$ decrease by one unit;
 this type of interference can be also represented as an additional source of leakage for the autoinducers, so that their
 overall leakage rate is $[\delta_A(N(t))+\delta_{IA} \eta_I(t)]A(t)$.
\end{itemize}
The model can be extended to include multiple interfering signals in a similar fashion. 
\section{Conclusions}
\label{conclusions}
In this position paper, we have explored how abstractions from communications,
networking and information theory can play a role in understanding and modeling bacterial
interactions. In particular, we have examined two forms of interactions in  bacterial systems: {\em electron transfer} and {\em quorum sensing}.
We have presented a stochastic queuing model of electron transfer for a single cell,
and we have showed how the proposed single-cell model can be extended to bacterial cables, by allowing for electron transfer  between neighboring cells.
We have performed a capacity analysis for a bacterial cable, demonstrating that such model is amenable to complexity reduction.
Moreover, we have provided a new queuing model for quorum sensing, which captures the dynamics of the quorum sensing signaling system, signal interactions,
and cell duplication within a stochastic framework.
We have shown that sufficient statistics can be identified, which allow a compact state space representation, and
we have provided preliminary simulation results.
Our investigation reveals that queuing models
effectively capture interactions in microbial communities and
 the inherent randomness exhibited by cell colonies in nature,
and they are amenable to complexity reduction using methods based on statistical physics
and wireless network design. Thus, they represent powerful
predictive tools, which will aid the design of systems exploiting bacterial capabilities.
\vspace{-7mm}
\section{Acknowledgments}
The authors would like to thank P. Silva for providing the experimental quorum sensing data.

\vspace{-4mm}
\bibliographystyle{IEEEtran}
\bibliography{IEEEabrv,Bib/RefsFIN} 

\begin{thebibliography}{10}
\providecommand{\url}[1]{#1}
\csname url@samestyle\endcsname
\providecommand{\newblock}{\relax}
\providecommand{\bibinfo}[2]{#2}
\providecommand{\BIBentrySTDinterwordspacing}{\spaceskip=0pt\relax}
\providecommand{\BIBentryALTinterwordstretchfactor}{4}
\providecommand{\BIBentryALTinterwordspacing}{\spaceskip=\fontdimen2\font plus
\BIBentryALTinterwordstretchfactor\fontdimen3\font minus
  \fontdimen4\font\relax}
\providecommand{\BIBforeignlanguage}[2]{{%
\expandafter\ifx\csname l@#1\endcsname\relax
\typeout{** WARNING: IEEEtran.bst: No hyphenation pattern has been}%
\typeout{** loaded for the language `#1'. Using the pattern for}%
\typeout{** the default language instead.}%
\else
\language=\csname l@#1\endcsname
\fi
#2}}
\providecommand{\BIBdecl}{\relax}
\BIBdecl

\bibitem{Kaufman}
A.~J. Kaufman, ``Early earth: Cyanobacteria at work,'' \emph{Nature
  Geoscience}, vol.~7, no.~4, pp. 253--254, 04 2014.

\bibitem{hogan2010bacteria}
C.~M. Hogan, ``Bacteria,'' \emph{Encyclopedia of Earth. Washington DC: National
  Council for Science and the Environment}, 2010.

\bibitem{KJB}
G.~A. Kowalchuk, S.~E. Jones, and L.~L. Blackall, ``Commentary: Microbes
  orchestrate life on earth.'' \emph{ISME Journal}, vol.~2, pp. 795--796, 2008.

\bibitem{Boedicker}
J.~Boedicker, M.~Vincent, and R.~Ismagilov, ``{Microfluidic Confinement of
  Single Cells of Bacteria in Small Volumes Initiates High-Density Behavior of
  Quorum Sensing and Growth and Reveals Its Variability},'' \emph{Angewandte
  Chemie-International Edition}, vol.~48, pp. 5908--5911, 2009.

\bibitem{Reguera}
G.~Reguera, K.~D. McCarthy, T.~Mehta, J.~S. Nicoll, M.~T. Tuominen, and D.~R.
  Lovley, ``{Extracellular electron transfer via microbial nanowires},''
  \emph{Nature}, vol. 435(7045), pp. 1098--1101, 2005.

\bibitem{Pfeffer}
C.~Pfeffer \emph{et~al.}, ``{Filamentous bacteria transport electrons over
  centimetre distances},'' \emph{Nature}, vol. 491(7423), pp. 218--221, 2012.

\bibitem{nakano2013molecular}
T.~Nakano, A.~W. Eckford, and T.~Haraguchi, \emph{Molecular
  communication}.\hskip 1em plus 0.5em minus 0.4em\relax Cambridge University
  Press, 2013.

\bibitem{mian2011communication}
I.~Mian and C.~Rose, ``Communication theory and multicellular biology,''
  \emph{Integrative Biology}, vol.~3, no.~4, pp. 350--367, 2011.

\bibitem{Naggar}
M.~Y. El-Naggar and S.~E. Finkel, ``{Live Wires: Electrical Signaling Between
  Bacteria},'' \emph{The Scientist}, vol.~27, no.~5, pp. 38--43, 2013.

\bibitem{Kato}
S.~Kato, K.~Hashimoto, and K.~Watanabe, ``{Microbial interspecies electron
  transfer via electric currents through conductive minerals},''
  \emph{Proceedings of the National Academy of Sciences}, vol. 109, no.~25, pp.
  10\,042--10\,046, June 2012.

\bibitem{silverman1955binary}
R.~Silverman \emph{et~al.}, ``On binary channels and their cascades,''
  \emph{IRE Transactions on Information Theory}, vol.~1, no.~3, pp. 19--27,
  1955.

\bibitem{simon}
M.~Simon, ``On the capacity of a cascade of identical discrete memoryless
  nonsingular channels,'' \emph{IEEE Transactions on Information Theory},
  vol.~16, no.~1, pp. 100--102, Jan 1970.

\bibitem{Pirbadian}
S.~Pirbadian and M.~Y. El-Naggar, ``{Multistep hopping and extracellular charge
  transfer in microbial redox chains},'' \emph{Physical Chemistry Chemical
  Physics}, vol.~14, pp. {13\,802--13\,808}, 2012.

\bibitem{coverthomas}
T.~M. Cover and J.~A. Thomas, \emph{{Elements of Information Theory}}.\hskip
  1em plus 0.5em minus 0.4em\relax New York, NY, USA: Wiley-Interscience, 1991.

\bibitem{einolghozati2013relaying}
A.~Einolghozati, M.~Sardari, and F.~Fekri, ``Relaying in diffusion-based
  molecular communication,'' in \emph{IEEE International Symposium on
  Information Theory Proceedings (ISIT)}.\hskip 1em plus 0.5em minus
  0.4em\relax IEEE, 2013, pp. 1844--1848.

\bibitem{Bassler}
M.~B. Miller and B.~L. Bassler, ``Quorum sensing in bacteria,'' \emph{Annual
  Review of Microbiology}, vol.~55, no.~1, pp. 165--199, 2001.

\bibitem{Visick15082005}
K.~L. Visick and C.~Fuqua, ``{Decoding Microbial Chatter: Cell-Cell
  Communication in Bacteria},'' \emph{Journal of Bacteriology}, vol. 187,
  no.~16, pp. 5507--5519, 2005.

\bibitem{Nealson70}
K.~Nealson, T.~Platt, and J.~Hastings, ``{Cellular Control of Synthesis and
  Activity of Bacterial Luminescent System},'' \emph{Journal of Bacteriology},
  vol. 104, no.~1, pp. 313--322, 1970.

\bibitem{kuran2012interference}
M.~{\c{S}}. Kuran, H.~B. Yilmaz, T.~Tugcu, and I.~F. Akyildiz, ``Interference
  effects on modulation techniques in diffusion based nanonetworks,''
  \emph{Nano Communication Networks}, vol.~3, no.~1, pp. 65--73, 2012.

\bibitem{arjmandi2013diffusion}
H.~Arjmandi, A.~Gohari, M.~N. Kenari, and F.~Bateni, ``Diffusion-based
  nanonetworking: A new modulation technique and performance analysis,''
  \emph{IEEE Communications Letters}, vol.~17, no.~4, pp. 645--648, 2013.

\bibitem{mosayebi2014receivers}
R.~Mosayebi, H.~Arjmandi, A.~Gohari, M.~Nasiri-Kenari, and U.~Mitra,
  ``Receivers for diffusion-based molecular communication: Exploiting memory
  and sampling rate,'' \emph{IEEE Journal on Selected Areas in Communications},
  vol.~32, no.~12, pp. 2368--2380, 2014.

\bibitem{noel2014optimal}
A.~Noel, K.~C. Cheung, and R.~Schober, ``Optimal receiver design for diffusive
  molecular communication with flow and additive noise,'' \emph{IEEE
  Transactions on NanoBioscience}, vol.~13, no.~3, pp. 350--362, 2014.

\bibitem{srinivas2012molecular}
K.~Srinivas, A.~W. Eckford, and R.~S. Adve, ``Molecular communication in fluid
  media: The additive inverse gaussian noise channel,'' \emph{IEEE Transactions
  on Information Theory}, vol.~58, no.~7, pp. 4678--4692, 2012.

\bibitem{li2014capacity}
H.~Li, S.~M. Moser, and D.~Guo, ``Capacity of the memoryless additive inverse
  gaussian noise channel,'' \emph{IEEE Journal on Selected Areas in
  Communications}, vol.~32, no.~12, pp. 2315--2329, 2014.

\bibitem{einolghozati2013design}
A.~Einolghozati, M.~Sardari, and F.~Fekri, ``{Design and analysis of wireless
  communication systems using diffusion-based molecular communication among
  bacteria},'' \emph{IEEE Transactions on Wireless Communications}, vol.~12,
  no.~12, pp. 6096--6105, 2013.

\bibitem{pierobon2013capacity}
M.~Pierobon and I.~F. Akyildiz, ``Capacity of a diffusion-based molecular
  communication system with channel memory and molecular noise,'' \emph{IEEE
  Transactions on Information Theory}, vol.~59, no.~2, pp. 942--954, 2013.

\bibitem{shannon1961two}
C.~E. Shannon \emph{et~al.}, ``Two-way communication channels,'' in \emph{Proc.
  4th Berkeley Symp. Math. Stat. Prob}, vol.~1, 1961, pp. 611--644.

\bibitem{van1971discrete}
E.~C. Van Der~Meulen, ``The discrete memoryless channel with two senders and
  one receiver,'' in \emph{Proc. IEEE Int. Symp. Information Theory (ISIT)},
  1971, p.~78.

\bibitem{liao1972multiple}
H.~H.-J. Liao, ``Multiple access channels,'' DTIC Document, Tech. Rep., 1972.

\bibitem{ahlswede1974capacity}
R.~Ahlswede, ``The capacity region of a channel with two senders and two
  receivers,'' \emph{The Annals of Probability}, pp. 805--814, 1974.

\bibitem{cover1972broadcast}
T.~M. Cover, ``Broadcast channels,'' \emph{IEEE Transactions on Information
  Theory}, vol.~18, no.~1, pp. 2--14, 1972.

\bibitem{cover1998comments}
------, ``Comments on broadcast channels,'' \emph{IEEE Transactions on
  information theory}, vol.~44, no.~6, pp. 2524--2530, 1998.

\bibitem{einolghozati2011consensus}
A.~Einolghozati, M.~Sardari, A.~Beirami, and F.~Fekri, ``Consensus problem
  under diffusion-based molecular communication,'' in \emph{45th Annual
  Conference on Information Sciences and Systems (CISS)}.\hskip 1em plus 0.5em
  minus 0.4em\relax IEEE, 2011, pp. 1--6.

\bibitem{Vicsek1995}
T.~Vicsek, A.~Czir\'ok, E.~Ben-Jacob, I.~Cohen, and O.~Shochet, ``Novel type of
  phase transition in a system of self-driven particles,'' \emph{Phys. Rev.
  Lett.}, vol.~75, pp. 1226--1229, Aug 1995.

\bibitem{Vicsek2012}
T.~Vicsek and A.~Zafeiris, ``Collective motion,'' \emph{Physics Reports}, vol.
  517, pp. 71--140, Aug. 2012.

\bibitem{Song2014}
H.-S. Song, W.~R. Cannon, A.~S. Beliaev, and A.~Konopka, ``Mathematical
  modeling of microbial community dynamics: A methodological review,''
  \emph{Processes}, vol.~2, no.~4, pp. 711--752, 2014.

\bibitem{Mina2012}
P.~Mina, M.~di~Bernardo, N.~J. Savery, and K.~Tsaneva-Atanasova, ``Modelling
  emergence of oscillations in communicating bacteria: a structured approach
  from one to many cells,'' \emph{Journal of The Royal Society Interface},
  vol.~10, no.~78, 2012.

\bibitem{baker2010microscopic}
R.~E. Baker, C.~A. Yates, and R.~Erban, ``From microscopic to macroscopic
  descriptions of cell migration on growing domains,'' \emph{Bulletin of
  mathematical biology}, vol.~72, no.~3, pp. 719--762, 2010.

\bibitem{Klapper2010}
I.~Klapper and J.~Dockery, ``Mathematical description of microbial biofilms,''
  \emph{SIAM Review}, vol.~52, no.~2, pp. 221--265, 2010.

\bibitem{Hammond2013}
J.~F. Hammond, E.~J. Stewart, J.~G. Younger, M.~J. Solomon, and D.~M. Bortz,
  ``Spatially heterogeneous biofilm simulations using an immersed boundary
  method with lagrangian nodes defined by bacterial locations,'' \emph{CoRR},
  vol. abs/1302.3663, 2013.

\bibitem{Friedman2014}
A.~Friedman, B.~Hu, and C.~Xue, ``\BIBforeignlanguage{English}{On a multiphase
  multicomponent model of biofilm growth},''
  \emph{\BIBforeignlanguage{English}{Archive for Rational Mechanics and
  Analysis}}, vol. 211, no.~1, pp. 257--300, 2014.

\bibitem{yates2011rsp}
C.~A. Yates, R.~Baker, R.~Erban, and P.~Maini, ``Refining self-propelled
  particle models for collective behaviour,'' \emph{Canadian Applied Maths
  Quarterly (CAMQ)}, vol.~18, no.~3, 2011.

\bibitem{waltermann2011information}
C.~Waltermann and E.~Klipp, ``Information theory based approaches to cellular
  signaling,'' \emph{Biochimica et Biophysica Acta (BBA)-General Subjects},
  vol. 1810, no.~10, pp. 924--932, 2011.

\bibitem{mehta2009information}
P.~Mehta, S.~Goyal, T.~Long, B.~L. Bassler, and N.~S. Wingreen, ``Information
  processing and signal integration in bacterial quorum sensing,''
  \emph{Molecular systems biology}, vol.~5, no.~1, p. 325, 2009.

\bibitem{cheong2011information}
R.~Cheong, A.~Rhee, C.~J. Wang, I.~Nemenman, and A.~Levchenko, ``Information
  transduction capacity of noisy biochemical signaling networks,''
  \emph{science}, vol. 334, no. 6054, pp. 354--358, 2011.

\bibitem{perez2011noise}
P.~D. P{\'e}rez, J.~T. Weiss, and S.~J. Hagen, ``Noise and crosstalk in two
  quorum-sensing inputs of vibrio fischeri,'' \emph{BMC systems biology},
  vol.~5, no.~1, p. 153, 2011.

\bibitem{Mitra}
U.~Mitra, N.~Michelusi, S.~Pirbadian, H.~Koorehdavoudi, M.~El-Naggar, and
  P.~Bogdan, ``{Queueing theory as a modeling tool for bacterial interaction:
  Implications for microbial fuel cells},'' in \emph{International Conference
  on Computing, Networking and Communications (ICNC)}, Feb 2015, pp. 658--662.

\bibitem{Levorato}
M.~Levorato, S.~Narang, U.~Mitra, and A.~Ortega, ``{Reduced dimension policy
  iteration for wireless network control via multiscale analysis},'' in
  \emph{IEEE Global Communications Conference (GLOBECOM)}, Dec 2012, pp.
  3886--3892.

\bibitem{rabaey2010microbial}
K.~Rabaey and R.~A. Rozendal, ``Microbial electrosynthesis -- revisiting the
  electrical route for microbial production,'' \emph{Nature Reviews
  Microbiology}, vol.~8, no.~10, pp. 706--716, 2010.

\bibitem{logan2009exoelectrogenic}
B.~E. Logan, ``Exoelectrogenic bacteria that power microbial fuel cells,''
  \emph{Nature Reviews Microbiology}, vol.~7, no.~5, pp. 375--381, 2009.

\bibitem{Kim2}
H.~J. Kim, H.~S. Park, M.~Hyun, I.~S. Chang, M.~Kim, and B.~Kim, ``{A
  mediator-less microbial fuel cell using a metal reducing bacterium,
  Shewanella putrefaciens},'' \emph{Enzyme and Microbial Technology}, vol.~30,
  no.~2, pp. 145--152, 2002.

\bibitem{mclean2010quantification}
J.~S. McLean, G.~Wanger, Y.~A. Gorby, M.~Wainstein, J.~McQuaid, S.~Ishii,
  O.~Bretschger, H.~Beyenal, and K.~H. Nealson, ``Quantification of electron
  transfer rates to a solid phase electron acceptor through the stages of
  biofilm formation from single cells to multicellular communities,''
  \emph{Environmental science \& technology}, vol.~44, no.~7, pp. 2721--2727,
  2010.

\bibitem{Gambello}
M.~Gambello and B.~Iglewski, ``{Cloning and characterization of the
  \emph{Pseudomonas aeruginosa} lasR gene, a transcriptional activator of
  elastase expression},'' \emph{Journal of Bacteriology}, vol. 173, pp.
  3000--3009, 1991.

\bibitem{Singh}
P.~Singh, A.~Schaefer, M.~Parsek, T.~Moninger, M.~Welsh, and E.~Greenberg,
  ``{Quorum-sensing signals indicate that cystic fibrosis lungs are infected
  with bacterial biofilms},'' \emph{Nature}, vol. 407, pp. 762--764, Oct. 2000.

\bibitem{Erickson}
D.~Erickson, R.~Endersby, A.~Kirkham, K.~Stuber, D.~Vollman, H.~Rabin,
  I.~Mitchell, and D.~Storey, ``{\emph{Pseudomonas aeruginosa} quorum-sensing
  systems may control virulence factor expression in the lungs of patients with
  cystic fibrosis},'' \emph{Infection and Immunity}, vol.~70, pp. 1783--1790,
  2002.

\bibitem{Rumbaugh}
K.~Rumbaugh, J.~Griswold, B.~Iglewski, and A.~Hamood, ``{Contribution of quorum
  sensing to the virulence of \emph{Pseudomonas aeruginosa} in burn wound
  infections},'' \emph{Infection and Immunity}, vol.~67, pp. 5854--5862, 1999.

\bibitem{Rasmussen}
T.~Rasmussen and M.~Givskov, ``{Quorum-sensing inhibitors as anti-pathogenic
  drugs},'' \emph{International Journal of Medical Microbiology}, vol. 296, pp.
  149--161, 2006.

\bibitem{Allen}
R.~Allen, R.~Popat, S.~Diggle, and S.~Brown, ``{Targeting virulence: can we
  make evolution-proof drugs?}'' \emph{Nature Reviews Microbiology}, vol.~12,
  pp. 300--308, 2014.

\bibitem{Kalia}
V.~Kalia, ``{Quorum sensing inhibitors: An overview},'' \emph{Biotechnology
  Advances}, vol.~31, pp. 224--245, 2013.

\bibitem{Lane}
N.~Lane, ``{Why Are Cells Powered by Proton Gradients?}'' \emph{Nature
  Education}, vol. 3(9):18, 2010.

\bibitem{JSACmiche}
N.~Michelusi, S.~Pirbadian, M.~El-Naggar, and U.~Mitra, ``{A Stochastic Model
  for Electron Transfer in Bacterial Cables},'' \emph{IEEE Journal on Selected
  Areas in Communications}, vol.~32, no.~12, pp. 2402--2416, Dec. 2014.

\bibitem{Ozalp}
V.~Ozalp, P.~T.R., N.~L.J., and O.~L.F., ``{Time-resolved measurements of
  intracellular ATP in the yeast Saccharomyces cerevisiae using a new type of
  nanobiosensor},'' \emph{Journal of Biological Chemistry}, pp.
  37\,579--37\,588, Nov. 2010.

\bibitem{smith2003foundations}
W.~Smith and J.~Hashemi, \emph{Foundations of Materials Science and
  Engineering}, ser. McGraw-Hill series in materials science and
  engineering.\hskip 1em plus 0.5em minus 0.4em\relax McGraw-Hill, 2003.

\bibitem{MicheIT}
N.~Michelusi and U.~Mitra, ``{Capacity of Electron-Based Communication Over
  Bacterial Cables: The Full-CSI Case},'' \emph{IEEE Transactions on Molecular,
  Biological and Multi-Scale Communications}, vol.~1, no.~1, pp. 62--75, March
  2015.

\bibitem{Chen}
J.~Chen and T.~Berger, ``{The capacity of finite-State Markov Channels with
  feedback},'' \emph{IEEE Transactions on Information Theory}, vol.~51, no.~3,
  pp. 780--798, March 2005.

\bibitem{Bertsekas2005}
D.~Bertsekas, \emph{Dynamic programming and optimal control}.\hskip 1em plus
  0.5em minus 0.4em\relax Athena Scientific, Belmont, Massachusetts, 2005.

\bibitem{Stauff}
D.~L. Stauff and B.~L. Bassler, ``{Quorum Sensing in \emph{Chromobacterium
  violaceum}: DNA Recognition and Gene Regulation by the CviR Receptor},''
  \emph{Journal of Bacteriology}, vol. 193, no.~15, pp. 3871--3878, Aug. 2011.

\bibitem{Stewart}
E.~Stewart, R.~Madden, G.~Paul, and F.~Taddei, ``{Aging and death in an
  organism that reproduces by morphologically symmetric division},'' \emph{PLoS
  Biology}, vol.~3, no.~2, Feb. 2005.

\bibitem{Dilanji}
G.~Dilanji, J.~Langebrake, P.~De~Leenheer, and S.~Hagen, ``{Quorum Activation
  at a Distance: Spatiotemporal Patterns of Gene Regulation from Diffusion of
  an Autoinducer Signal},'' \emph{Journal of the American Chemical Society},
  vol. 134, pp. 5618--5626, 2012.

\bibitem{Pai}
A.~Pai, J.~Srimani, Y.~Tanouchi, and L.~You, ``{Generic metric to quantify
  quorum sensing activation dynamics},'' \emph{ACS synthetic biology}, vol.~3,
  pp. 220--227, 2014.

\bibitem{Kastrup}
C.~J. Kastrup, F.~Shen, and R.~F. Ismagilov, ``{Response to Shape Emerges in a
  Complex Biochemical Network and Its Simple Chemical Analogue},''
  \emph{Angewandte Chemie International Edition}, vol.~46, no.~20, pp.
  3660--3662, 2007.

\bibitem{Gray}
K.~Gray and J.~R. Garey, ``{The evolution of bacterial LuxI and LuxR quorum
  sensing regulators},'' \emph{Microbiology}, vol. 147, no.~8, pp. 2379--2387,
  Aug. 2001.

\bibitem{Teng2}
S.-W. Teng, Y.~Wang, K.~Tu, T.~Long, P.~Mehta, N.~S. Wingreen, B.~L. Bassler,
  and N.~Ong, ``{Measurement of the Copy Number of the Master Quorum-Sensing
  Regulator of a Bacterial Cell},'' \emph{Biophysical Journal}, vol.~98, no.~9,
  pp. 2024--2031, May 2010.

\bibitem{Surette}
M.~G. Surette, M.~B. Miller, and B.~L. Bassler, ``{Quorum sensing in
  \emph{Escherichia coli}, \emph{Salmonella typhimurium}, and \emph{Vibrio
  harveyi}: A new family of genes responsible for autoinducer production},''
  \emph{PNAS}, vol.~96, no.~4, pp. 1639--1644, 1999.

\bibitem{PerezVelazquez}
J.~Perez-Velazquez, B.~Quinones, B.~A. Hense, and C.~Kuttler, ``{A mathematical
  model to investigate quorum sensing regulation and its heterogeneity in
  \emph{Pseudomonas syringae} on leaves},'' \emph{Ecological Complexity},
  vol.~21, pp. 128--141, 2015.

\bibitem{Carnes}
E.~Carnes, D.~Lopez, N.~Donegan, A.~Cheung, H.~Gresham, G.~Timmins, and
  C.~Brinker, ``{Confinement-Induced Quorum Sensing of Individual
  Staphylococcus aureus Bacteria},'' \emph{Nature chemical biology}, vol.~6,
  pp. 41--45, 2010.

\bibitem{Boedicker2}
J.~Boedicker, H.~Gargia, and R.~Phillips, ``{Theoretical and Experimental
  Dissection of DNA Loop-Mediated Repression},'' \emph{Physical Review
  Letters}, vol. 110, no.~1, 2013.

\bibitem{McLean}
R.~McLean, L.~Pierson~III, and C.~Fuqua, ``{A simple screening protocol for the
  identification of quorum signal antagonists},'' \emph{Journal of
  Microbiological Methods}, vol.~58, pp. 351--360, 2004.

\bibitem{Kirisits}
M.~Kirisits, J.~Margolis, B.~Purevdorj-Gage, B.~Vaughan, D.~Chopp, P.~Stoodley,
  and M.~Parsek, ``{Influence of the hydrodynamic environment on quorum sensing
  in \emph{Pseudomonas aeruginosa} biofilms},'' \emph{Journal of bacteriology},
  vol. 189, pp. 8357--8360, 2007.

\bibitem{Chandler}
J.~R. Chandler, S.~Heilmann, J.~E. Mittler, and E.~P. Greenberg,
  ``{Acyl-homoserine lactone-dependent eavesdropping promotes competition in a
  laboratory co-culture model},'' \emph{ISME Journal}, vol.~6, no.~12, pp.
  2219--2228, 12 2012.

\bibitem{Smith}
C.~Smith, H.~Song, and L.~You, ``{Signal discrimination by differential
  regulation of protein stability in quorum sensing},'' \emph{Journal of
  Molecular Biology}, no. 382, pp. 1290--1297, Oct. 2008.

\bibitem{Bionumbers}
\BIBentryALTinterwordspacing
Bionumbers. [Online]. Available: \url{http://bionumbers.hms.harvard.edu}
\BIBentrySTDinterwordspacing

\bibitem{Gillespie_1977}
D.~T. Gillespie, ``{Exact stochastic simulation of coupled chemical
  reactions},'' \emph{J. Physical Chemistry}, vol.~81, no.~25, pp. 2340--2361,
  Dec. 1977.

\bibitem{Kastrup2}
C.~J. Kastrup, J.~Boedicker, A.~Pomerantsev, M.~Moayeri, Y.~Bian, R.~Pompano,
  T.~Kline, P.~Sylvester, F.~Shen, S.~Leppla, W.~Tang, and R.~Ismagilov,
  ``{Spatial localization of bacteria controls coagulation of human blood by
  'quorum acting'},'' \emph{Nature Chemical Biology}, vol.~4, no.~12, pp.
  742--750, 2008.

\bibitem{Prindle}
A.~Prindle, P.~Samayoa, I.~Razinkov, T.~Danino, L.~Tsimring, and J.~Hasty, ``{A
  sensing array of radically coupled genetic "biopixels"},'' \emph{Nature},
  vol. 481, no. 7379, pp. 39--44, Dec. 2011.

\bibitem{Jimenez}
P.~Jimenez, G.~Koch, J.~A. Thompson, K.~B. Xavier, R.~H. Cool, and W.~J. Quax,
  ``{The Multiple Signaling Systems Regulating Virulence in \emph{Pseudomonas
  aeruginosa}},'' \emph{Microbiology and Molecular Biology Reviews}, vol.~76,
  no.~1, pp. 46--65, March 2012.

\bibitem{Lee}
J.~Lee, J.~Wu, Y.~Deng, J.~Wang, C.~Wang, J.~Wang, C.~Chang, Y.~Dong,
  P.~Williams, and L.~Zhang, ``{A cell-cell communication signal integrates
  quorum sensing and stress response},'' \emph{Nature Chemical Biology},
  vol.~9, no.~5, pp. 339--343, Mar. 2013.

\bibitem{Teng}
S.~Teng, J.~Schaffer, K.~Tu, P.~Mehta, W.~Lu, N.~Ong, B.~Bassler, and
  N.~Wingreen, ``{Active regulation of receptor ratios controls integration of
  quorum-sensing signals in \emph{Vibrio harveyi}},'' \emph{Molecular Systems
  Biology}, vol.~7, no. 491, May 2011.

\bibitem{Kimura}
N.~Kimura, ``{Metagenomic approaches to understanding phylogenetic diversity in
  quorum sensing},'' \emph{Virulence}, vol.~5, no.~3, pp. 433--442, Apr. 2014.

\bibitem{Rampioni201460}
G.~Rampioni, L.~Leoni, and P.~Williams, ``The art of antibacterial warfare:
  Deception through interference with quorum sensing-mediated communication,''
  \emph{Bioorganic Chemistry}, vol.~55, no.~0, pp. 60 -- 68, 2014, bio-Organic
  Chemistry of Antibacterial Drug Discovery.

\bibitem{Chan}
K.-G. Chan, S.~Atkinson, K.~Mathee, C.-K. Sam, S.~R. Chhabra, M.~Camara, C.-L.
  Koh, and P.~Williams, ``{Characterization of N-acylhomoserine
  lactone-degrading bacteria associated with the Zingiber officinale (ginger)
  rhizosphere: Co-existence of quorum quenching and quorum sensing in
  \emph{Acinetobacter} and \emph{Burkholderia}},'' \emph{BMC Microbiology},
  vol.~11, no.~51, 2011.

\bibitem{Gonzalez}
J.~E. Gonzalez and N.~D. Keshavan, ``{Messing with Bacterial Quorum Sensing},''
  \emph{Microbiology and Molecular Biology Reviews}, vol.~70, no.~4, p.
  859Ð875, Dec. 2006.

\bibitem{Williams}
P.~Williams, K.~Winzer, W.~Chan, and M.~Camara, ``{Look who's talking:
  communication and quorum sensing in the bacterial world},''
  \emph{Philosophical Transactions of the Royal Society B: Biological
  Sciences}, vol. 362, no. 1483, pp. 1119--1134, July 2007.

\bibitem{McClean}
K.~McClean, M.~Winson, L.~Fish, A.~Taylor, S.~Chhabra, M.~Camara, M.~Daykin,
  J.~Lamb, S.~Swift, B.~Bycroft, G.~Stewart, and P.~Williams, ``{Quorum sensing
  and Chromobacterium violaceum: exploitation of violacein production and
  inhibition for the detection of N-acylhomoserine lactones},''
  \emph{Microbiology}, vol. 143, no.~12, pp. 3703--3711, Dec. 1997.

\end{thebibliography}
\begin{IEEEbiography}[{\includegraphics[width=1in,height=1.25in,clip,keepaspectratio]{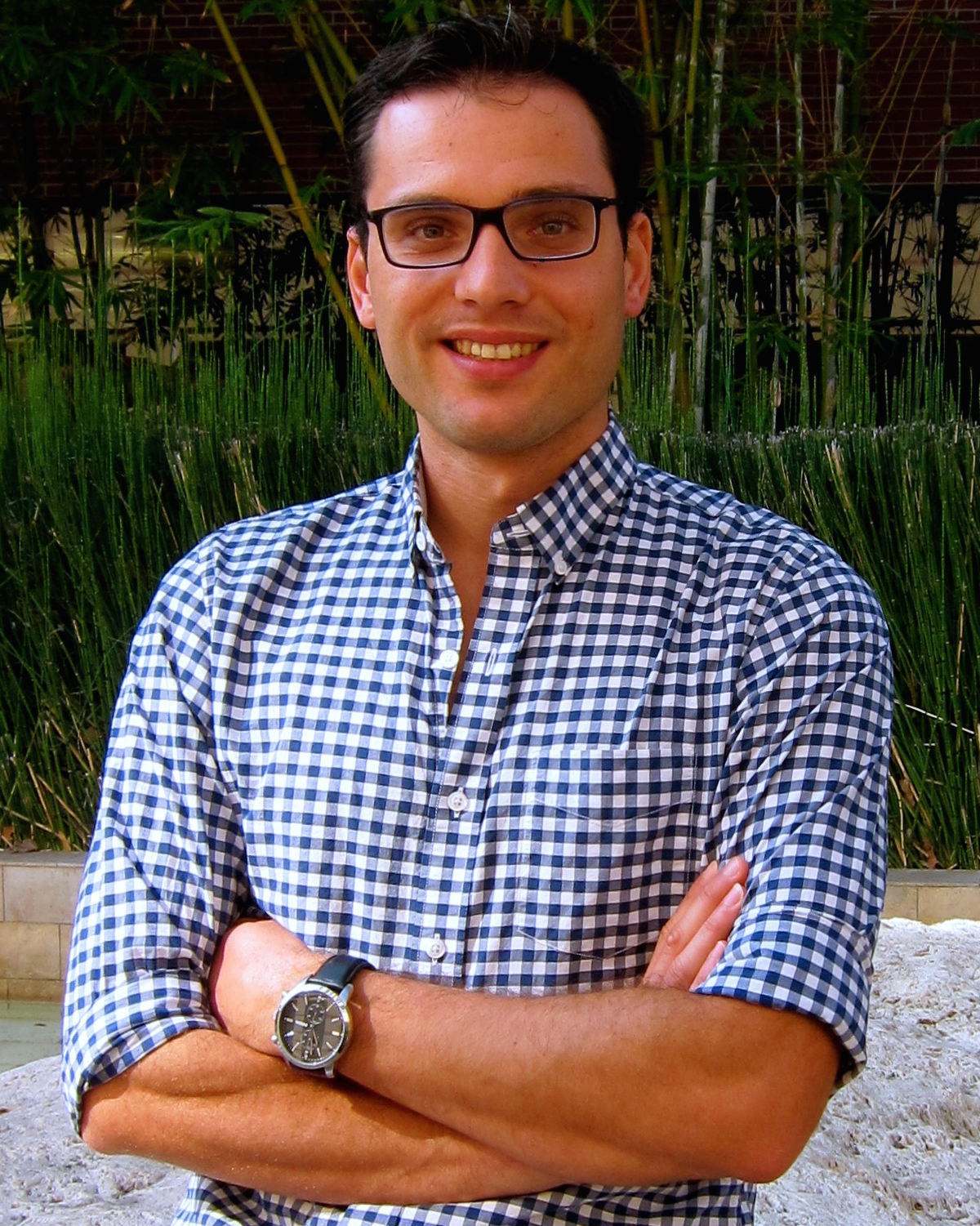}}]
{Nicol\`{o}~Michelusi} (S'09, M'13) received the B.Sc. (with honors), M.Sc.
 (with honors) and Ph.D. degrees from the University of Padova, Italy, in 2006, 2009 and 2013, respectively,
and the M.Sc. degree in Telecommunications Engineering from the Technical University of Denmark in 2009, as part of the T.I.M.E. double degree program.
In 2011, he was at the University
of Southern California, Los Angeles, USA,
and, in Fall 2012, at Aalborg University, Denmark, as a visiting research scholar.
He was a post-doctoral research fellow at the Ming Hsieh Department of Electrical Engineering, University of Southern California, USA, in 2013-2015.
He is currently an Assistant Professor in the Department of Electrical and Computer Engineering at Purdue University, IN, USA.
 His research interests lie in the areas of
wireless networks, stochastic optimization, distributed estimation and modeling of bacterial networks.
  Dr. Michelusi  serves as a reviewer for the IEEE Transactions on Communications, IEEE Transactions on Wireless Communications, IEEE Transactions on Information Theory, IEEE Transactions on Signal Processing, IEEE Journal on Selected Areas in Communications,
 and IEEE/ACM Transactions on Networking.
 \end{IEEEbiography}
 
 \begin{IEEEbiography}[{\includegraphics[width=1in,height=1.25in,clip,keepaspectratio]{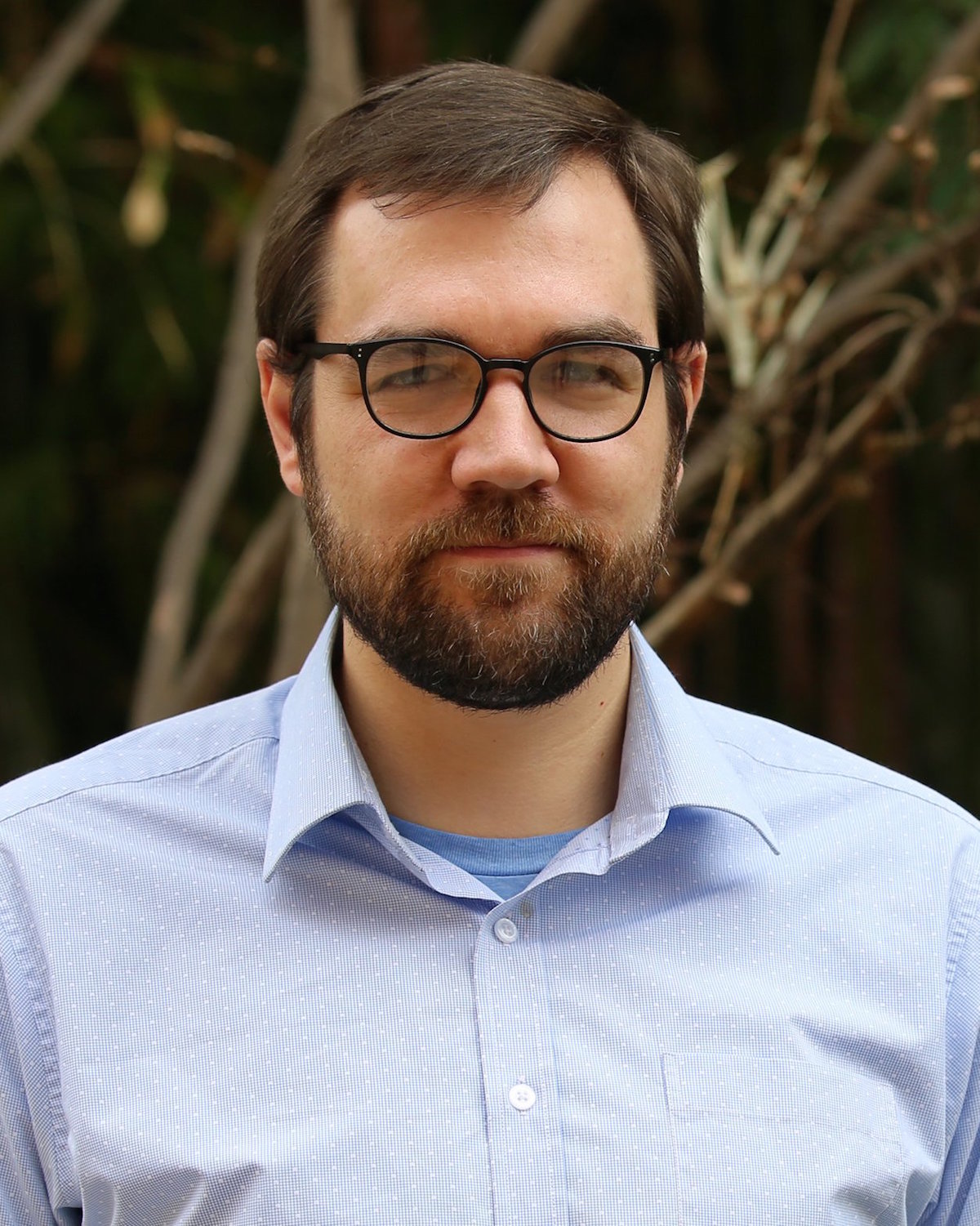}}]
 {James Boedicker}
received a S.B. from the Massachusetts Institute of
Technology, and M.A. and Ph.D. degrees from the University of Chicago.
He was a Postdoctoral Scholar at the California Institute of
Technology from 2010-2013 in the Department of Applied Physics.  He is
currently an Assistant Professor of Physics and Astronomy and the
Biological Sciences at the University of Southern California.  His
research develops experimental and theoretical tools to understand the
emergent properties of cellular networks, signal exchange in microbial
ecosystems, and the consequences of stochasticity in biological
systems.
\end{IEEEbiography}

\begin{IEEEbiography}[{\includegraphics[width=1in,height=1.25in,clip,keepaspectratio]{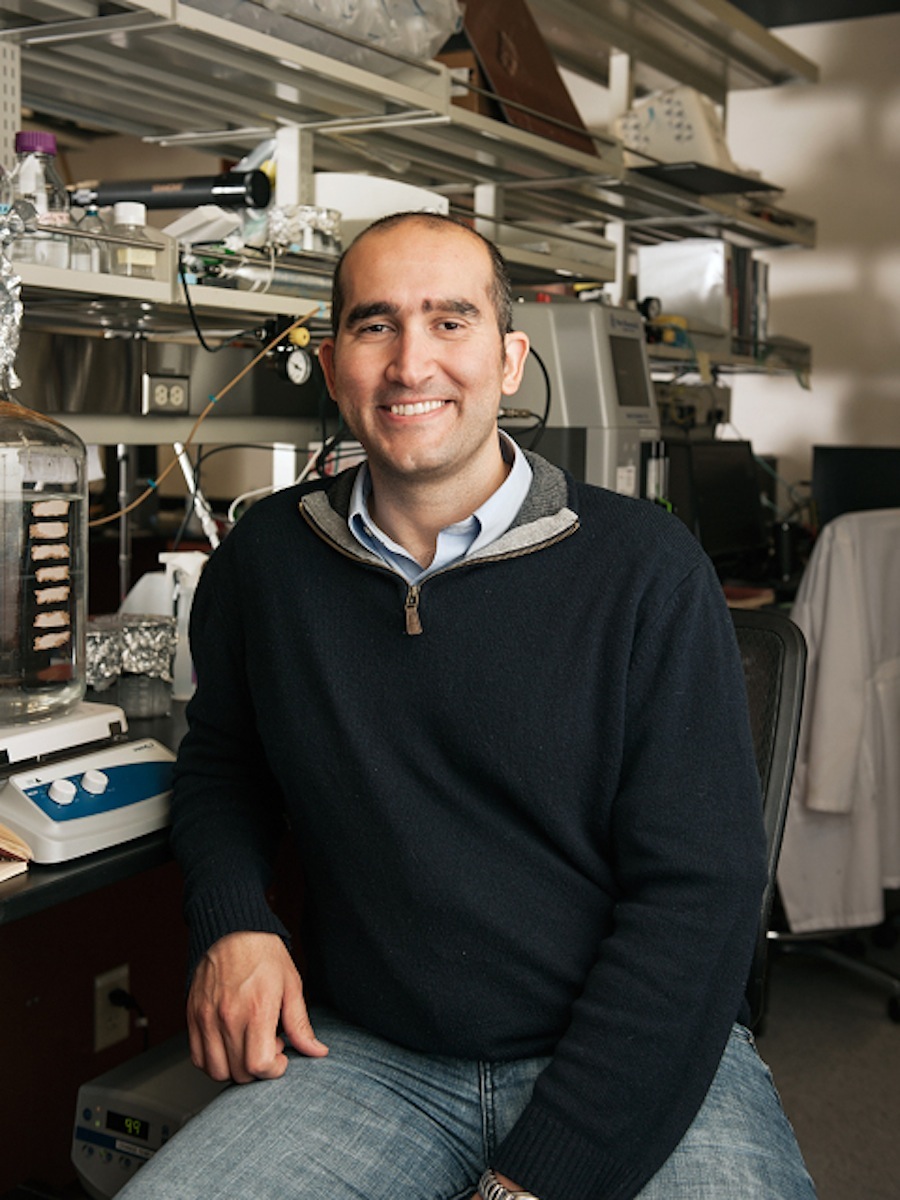}}]
{Moh El-Naggar}  is an assistant professor of physics at the University of Southern California's Dornsife College of Letters, Arts and Sciences. El-Naggar studies energy conversion and charge transmission at the interface between living cells and natural or synthetic surfaces. In 2012, he was named one of Popular ScienceÕs ÔBrilliant 10Õ, and in 2014 he was awarded the Presidential Early Career Award for Scientists and Engineers (PECASE). El-NaggarÕs work has important implications for cell physiology and astrobiology: it may lead to the development of new hybrid materials and renewable energy technologies that combine the exquisite biochemical control of nature with the synthetic building blocks of nanotechnology. El-Naggar earned his B.S. degree in mechanical engineering from Lehigh University (2001), and his Ph.D. in engineering and applied science from the California Institute of Technology (2007).
 \end{IEEEbiography}

\begin{IEEEbiographynophoto}{Urbashi Mitra} (F'07)
received the B.S. and the M.S. degrees from the University of California at Berkeley and her Ph.D. from Princeton University. After a six-year stint at the Ohio State University, she joined the Department of Electrical Engineering at the University of Southern California, Los Angeles, where she is currently a Dean's Professor. Dr. Mitra is a member of the IEEE Information Theory Society's Board of Governors (2002-2007, 2012-2017) and the IEEE Signal Processing Society's Technical Committee on Signal Processing for Communications and Networks (2012-2016). Dr. Mitra is a Fellow of the IEEE.  She is the inaugural Editor-in-Chief of the IEEE Transactions on Molecular, Biological and Multi-scale Communications. Dr. Mitra is the recipient of: INSIGHT Into Diversity 2015 Inspiring Women in STEM Award, a 2014-2015 IEEE Communications Society Distinguished Lecturer, a Tutorial Lecturer for the IEEE North American School on Information Theory, UC San Diego, August 2015, the 2012 Globecom Signal Processing for Communications Symposium Best Paper Award, a 2012 NAE Lillian Gilbreth Lectureship, an USC Center for Excellence in Research Fellowship (2010-2013), the 2009 USC VSoE Dean's Service Award, the 2009 DCOSS Applications \& Systems Best Paper Award,  a 2009 USC Remarkable Woman Award (Faculty) , a 2008 USC Mellon Mentoring Award (Faculty-to-Faculty), the Texas Instruments Visiting Professorship (Fall 2002, Rice University), 2001 Okawa Foundation Award, 2000 OSU College of Engineering Lumley Award for Research, 1997 OSU College of Engineering MacQuigg Award for Teaching, and a 1996 National Science Foundation (NSF) CAREER Award. Dr. Mitra currently serves on the IEEE Fourier Award for Signal Processing, the IEEE James H. Mulligan, Jr. Education Medal and the IEEE Paper Prize committees. She has been an Associate Editor for the following IEEE publications: Transactions on Signal Processing (2012-2015), Transactions on Information Theory (2007-2011), Journal of Oceanic Engineering (2006-2011), and Transactions on Communications (1996-2001). She has co-chaired: (technical program) 2014 IEEE International Symposium on Information Theory in Honolulu, HI, 2014 IEEE Information Theory Workshop in Hobart, Tasmania, IEEE 2012 International Conference on Signal Processing and Communications, Bangalore India, and  the IEEE  Communication Theory Symposium at ICC 2003 in Anchorage, AK;  and  was the general co-chair for the first ACM Workshop on Underwater Networks at Mobicom 2006, Los Angeles, CA Dr. Mitra was the Tutorials Chair for IEEE ISIT 2007 in Nice, France and the Finance Chair for IEEE ICASSP 2008 in Las Vegas, NV.  She served as co-Director of the Communication Sciences Institute at the University of Southern California from 2004-2007.  Her research interests are in: wireless communications, biological communication, underwater acoustic communications, communication and sensor networks, detection and estimation and the interface of communication, sensing and control.
  \end{IEEEbiographynophoto}

\end{document}